\documentclass[sigconf, nonacm, pdfa]{acmart}

\newcommand\vldbdoi{10.14778/3659437.3659449}
\newcommand\vldbpages{1939 - 1952}
\newcommand\vldbvolume{17}
\newcommand\vldbissue{8}
\newcommand\vldbyear{2024}
\newcommand\vldbauthors{\authors}
\newcommand\vldbtitle{\shorttitle} 
\newcommand\vldbavailabilityurl{https://github.com/SolidLao/GPTuner}
\newcommand\vldbpagestyle{empty} 

\usepackage{tcolorbox}
\usepackage{balance} 
\usepackage{booktabs} 
\usepackage{amsmath}
\usepackage{graphicx}

\newcommand{\system}{{\sc GPTuner }}

\usepackage[linesnumbered,ruled,vlined]{algorithm2e} 
\usepackage{epstopdf}
\usepackage{multicol}
\usepackage{url}
\usepackage{enumerate}
\usepackage[labelfont=bf]{caption}
\usepackage{xcolor}
\usepackage{color, colortbl}
\usepackage{hyperref}
\usepackage{algpseudocode}
\usepackage{tabularx}

\usepackage{subfig}
\usepackage{graphicx}

\usepackage{enumitem} 

\usepackage{multirow}
\usepackage{diagbox}
\usepackage{CJKutf8} 

\usepackage{boldline}

\usepackage{soul} 

\DeclareMathOperator*{\argmax}{arg\,max} 

\hypersetup{
  colorlinks   = true, 
  urlcolor     = blue, 
  linkcolor    = blue, 
  citecolor   = red 
}
\usepackage{pifont}

\definecolor{codegreen}{rgb}{0,0.6,0}
\definecolor{codegray}{rgb}{0.5,0.5,0.5}
\definecolor{codepurple}{rgb}{0.58,0,0.82}
\definecolor{backcolour}{rgb}{0.95,0.95,0.92}
\definecolor{lightgray}{gray}{0.9}

\begin{document}
\pagestyle{plain} 

\title{\textsc{GPTuner}: A Manual-Reading Database Tuning System via GPT-Guided Bayesian Optimization}


\author{Jiale Lao}
\orcid{0009-0003-1144-5152}
\affiliation{%
  \institution{Sichuan University}
}
\email{solidlao.jiale@gmail.com}

\author{Yibo Wang}
\orcid{0009-0005-1971-3398}
\affiliation{%
  \institution{Sichuan University}
}
\email{wangyibo.cs@gmail.com}

\author{Yufei Li}
\orcid{0009-0004-4285-5696}
\affiliation{%
  \institution{Sichuan University}
}
\email{liyufeievangeline@gmail.com}

\author{Jianping Wang}
\orcid{0009-0001-1893-4245}
\affiliation{%
  \institution{Northwest Normal University}
}
\email{2022222119@nwnu.edu.cn}

\author{Yunjia Zhang}
\orcid{0009-0001-7157-156X}
\affiliation{%
  \institution{University of Wisconsin-Madison}
}
\email{yunjia@cs.wisc.edu}

\author{Zhiyuan Cheng}
\orcid{0000-0001-7280-6079}
\affiliation{%
  \institution{Purdue University}
}
\email{cheng443@purdue.edu}

\author{Wanghu Chen}
\orcid{0000-0002-9233-7609}
\affiliation{%
  \institution{Northwest Normal University}
}
\email{chenwh@nwnu.edu.cn}

\author{Mingjie Tang}
\orcid{0000-0002-8893-4574}
\authornote{The corresponding author.}
\affiliation{%
  \institution{Sichuan University}
}
\email{tangrock@gmail.com}

\author{Jianguo Wang}
\orcid{0000-0002-3039-1175}
\affiliation{%
  \institution{Purdue University}
}
\email{csjgwang@purdue.edu}
 
\begin{abstract}
Modern database management systems (DBMS) expose hundreds of configurable knobs to control system behaviours. Determining the appropriate values for these knobs to improve DBMS performance is a long-standing problem in the database community. As there is an increasing number of knobs to tune and each knob could be in continuous or categorical values, manual tuning becomes impractical. Recently, automatic tuning systems using machine learning methods have shown great potentials. However, existing approaches still incur significant tuning costs or only yield sub-optimal performance. This is because they either ignore the extensive domain knowledge available (e.g., DBMS manuals and forum discussions) and only rely on the runtime feedback of benchmark evaluations to guide the optimization, or they utilize the domain knowledge in a limited way. Hence, we propose {\sc GPTuner}, a manual-reading database tuning system that leverages domain knowledge extensively and automatically to optimize search space and enhance the runtime feedback-based optimization process. Firstly, we develop a Large Language Model (LLM)-based pipeline to collect and refine heterogeneous knowledge, and propose a prompt ensemble algorithm to unify a structured view of the refined knowledge. Secondly, using the structured knowledge, we (1) design a workload-aware and training-free knob selection strategy, (2) develop a search space optimization technique considering the value range of each knob, and (3) propose a Coarse-to-Fine Bayesian Optimization Framework to explore the optimized space. Finally, we evaluate \system under different benchmarks (TPC-C and TPC-H), metrics (throughput and latency) as well as DBMS (PostgreSQL and MySQL). Compared to the state-of-the-art approaches, \system identifies better configurations in \textbf{16x} less time on average. Moreover, \system achieves up to \textbf{30\%} performance improvement (higher throughput or lower latency) over the \textbf{best-performing} alternative.

\end{abstract}

\maketitle

\pagestyle{\vldbpagestyle}
\begingroup\small\noindent\raggedright\textbf{PVLDB Reference Format:}\\
\vldbauthors. \vldbtitle. PVLDB, \vldbvolume(\vldbissue): \vldbpages, \vldbyear.\\
\href{https://doi.org/\vldbdoi}{doi:\vldbdoi}
\endgroup
\begingroup
\renewcommand\thefootnote{}\footnote{\noindent
This work is licensed under the Creative Commons BY-NC-ND 4.0 International License. Visit \url{https://creativecommons.org/licenses/by-nc-nd/4.0/} to view a copy of this license. For any use beyond those covered by this license, obtain permission by emailing \href{mailto:info@vldb.org}{info@vldb.org}. Copyright is held by the owner/author(s). Publication rights licensed to the VLDB Endowment. \\
\raggedright Proceedings of the VLDB Endowment, Vol. \vldbvolume, No. \vldbissue\ %
ISSN 2150-8097. \\
\href{https://doi.org/\vldbdoi}{doi:\vldbdoi} \\
}\addtocounter{footnote}{-1}\endgroup



\ifdefempty{\vldbavailabilityurl}{}{
\vspace{.3cm}
\begingroup\small\noindent\raggedright\textbf{PVLDB Artifact Availability:}\\
The source code, data, and/or other artifacts have been made available at \url{\vldbavailabilityurl}. 
\endgroup
}

%

\section{Introduction}\label{sec:intro}

Modern Database Management Systems (DBMS) expose hundreds of configurable parameters (i.e., knobs) to control their runtime behaviours \cite{pavlo2017self}. The selection of appropriate values for these knobs is crucial to improve DBMS performance, constituting a long-standing challenge within the database community \cite{10.14778/3450980.3450992}. Given the high dimensionality of the configuration space (e.g., PostgreSQL v14.9 has 346 knobs), and the inherent heterogeneity of these knobs (due to their continuous and categorical domains), database administrators (DBAs) encounter substantial difficulties in identifying promising configurations tailored to specific query workloads. The magnitude of this challenge becomes even more pronounced in the cloud environment, where the underlying physical configurations can significantly vary among distinct DBMS instances \cite{10.14778/3476311.3476411}.


To reduce the manual tuning efforts of DBAs, state-of-the-art approaches automate the knob tuning via Machine Learning (ML) techniques, including Bayesian Optimization~\cite{10.14778/1687627.1687767, 10.1145/3035918.3064029, 10.1145/3448016.3457291, 10.14778/3457390.3457404, 10.1145/3514221.3526176, 10.1145/3318464.3380591} and Reinforcement Learning~\cite{10.1145/3299869.3300085, 10.14778/3352063.3352129, 10.1145/3514221.3517882, 10.1145/3514221.3517843}. These ML-based tuning systems follow the main concept of ``trial and error'' to explore the configuration space iteratively, balancing between the exploration of unseen regions and the exploitation of known space. 

While these methods do possess the potential to reach well-performing knob configurations eventually, they still incur significant tuning costs~\cite{10.14778/3551793.3551844, 10.1145/3318464.3380591}. Previous studies~\cite{10.1145/3514221.3517882, 10.1145/3448016.3457291, 10.14778/3538598.3538604} have revealed that state-of-the-art systems still require hundreds to thousands iterations to reach an ideal configuration, with each iteration taking minutes or more to execute the target workload. Such high tuning costs stem from their inefficiency in handling: (1) \textit{the large number of knobs that requires tuning}, and (2) \textit{the wide search space of possible values for each knob}. For the first difficulty, most approaches either select a fixed subset of knobs \cite{10.1145/3514221.3517882, 10.14778/3551793.3551844, 10.14778/3457390.3457404, 10.14778/1687627.1687767, 10.1145/3299869.3300085, 10.1145/3127479.3128605}, sacrificing the flexibility to choose workload-relevant knobs, or execute workloads numerous times to identify important knobs \cite{10.14778/3538598.3538604, 10.1145/3035918.3064029, 10.5555/3488733.3488749}, which is resource-intensive. Regarding the second difficulty, most approaches use the default value ranges provided by DBMS vendors~\cite{10.14778/1687627.1687767, 10.1145/3035918.3064029, 10.1145/3448016.3457291, 10.14778/3457390.3457404, 10.1145/3514221.3526176, 10.1145/3318464.3380591, 10.1145/3299869.3300085, 10.14778/3352063.3352129, 10.1145/3514221.3517882}. However, the default value ranges are excessively broad for flexibility, which complicates the tuning process and introduces the risk of system crashes \cite{10.14778/3476311.3476411, 10.14778/3450980.3450992}.

In contrast to ML-based tuning approaches that adjust the DBMS solely based on performance statistics, human DBAs often rely on domain knowledge for tuning (e.g., DBMS manuals and discussions from DBMS forums). Unlike performance statistics, external domain knowledge directly reveals tuning hints, including the important knobs and typical value ranges for each knob. First, there are discussions on which knobs significantly impact DBMS performance. For instance, in web forums like \textit{Hacker News}, it is commonly mentioned that parallel knobs (e.g.,  ``\texttt{max\_parallel\_workers\_per\_gather}'') are crucial for OLAP workloads, while I/O-related knobs (e.g., ``\texttt{max\_wal\_size}'') are important for OLTP workloads \cite{hackernews}. Second, there are typical value ranges summarized for knobs. Table~\ref{table: guidance-improvement} provides two examples of extracting improved value ranges from natural language tuning guidance. For knob ``\texttt{shared\_buffers}'', instead of using the default value range [0.125 MB, 8192 GB], the improved value range can be [4 GB, 6.4 GB] since the guidance suggests setting the value between 25\% and 40\% of the RAM (on a machine with 16 GB of RAM). For knob ``\texttt{random\_page\_cost}'', we can try out value range [1.0, 2.0] rather than [$0, 1.79769 \times 10^{308}$] if the machine uses SSDs as disks. We can observe that these hints are highly valuable in reducing the search space of ML-based methods and thus expedite convergence and achieve better performance. 


\begin{table}[t]
\centering
\caption{Tuning Knowledge Utilization}
 
\footnotesize
\setlength{\tabcolsep}{4pt}
\renewcommand{\arraystretch}{1}
\begin{tabularx}{\columnwidth}{>{\bfseries}lXX} 
\hlineB{2}
Knob & \bfseries \texttt{shared\_buffers} & \bfseries \texttt{random\_page\_cost} \\
\hline
\rowcolor{lightgray}
Default Range & $[0.125 \text{MB},~8192~\text{GB}]$ & $[0,~1.79769 \times 10^{308}]$ \\
\hline
Guidance & ``\texttt{shared\_buffers}'' can be \textcolor{magenta}{25\%} of the RAM but no more than \textcolor{magenta}{40\%} \dots \cite{PostgreSQL2023} & ``\texttt{random\_page\_cost}'' can be \textcolor{magenta}{1.x} if disk has a speed similar to SSDs \dots \cite{PostgreSQLCONF} \\
\hline
DBA & The machine has a \textcolor{blue}{16 GB} RAM. Thus we can set ``\texttt{shared\_buffers}'' from \textcolor{blue}{16 GB $\times$ 25\% = 4 GB} to \textcolor{blue}{16 GB $\times$ 40\% = 6.4 GB}. & The machine uses SSDs as disks. Thus we can set ``\texttt{random\_page\_cost}'' to a value from \textcolor{blue}{1.0} to \textcolor{blue}{2.0}. \\
\hline
\rowcolor{lightgray}
Improved Range & \textcolor{red}{[4 GB, 6.4 GB]} & \textcolor{red}{[1.0, 2.0]} \\
\hlineB{2}
\end{tabularx}
\label{table: guidance-improvement}
 
\end{table}

%
%

\begin{figure}[tb]
    \centering
    \subfloat[TPC-H Benchmark]{
        \includegraphics[scale=0.25]{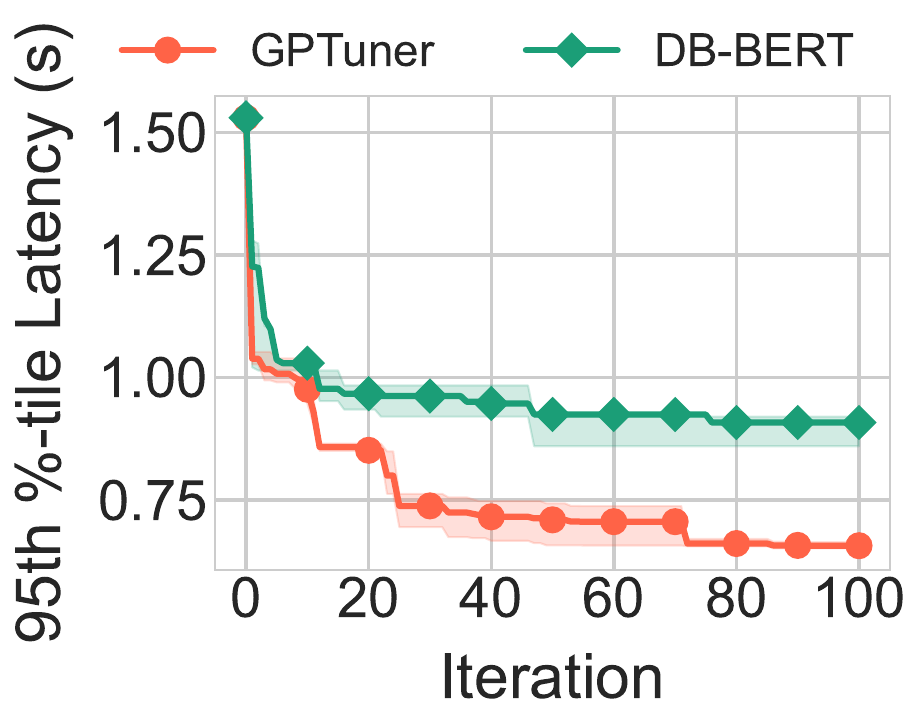}
        \label{fig: intro-compare-tpch}
    }
    \subfloat[TPC-C Benchmark]{
        \includegraphics[scale=0.25]{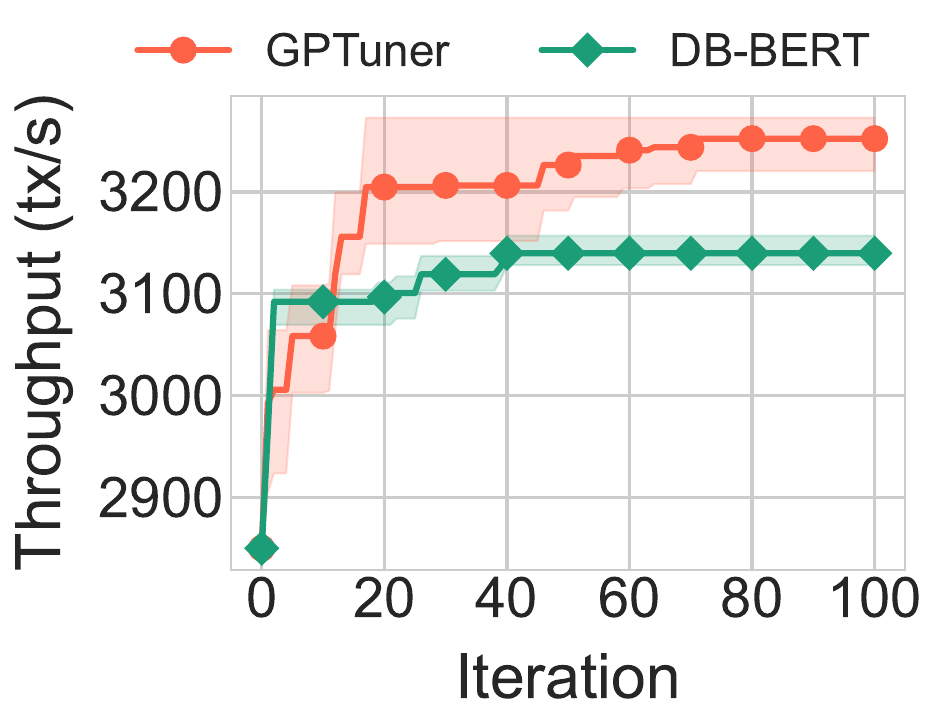}
        \label{fig: intro-compare-tpcc}
    }
     
    \caption{{\sc \bfseries GPTuner} vs. DB-BERT}
    \label{fig: intro-compare}
     
\end{figure}

Therefore, to mitigate the high tuning costs of ML-based techniques, it is desirable to design a knob-tuning system leveraging domain knowledge to enhance the optimization process. However, this is non-trivial for the following challenges. \textit{C1. It is challenging to unify a structured view of the heterogeneous domain knowledge while balancing a trade-off between cost and quality.} Domain knowledge typically comes in the form of DBMS manuals and discussions on the web forums. To leverage such knowledge in the ML-based techniques, we need to transfer it into a unified machine-readable format (i.e., structured data). However, preparing such a structured view involves a complex and lengthy workflow: data ingestion, data cleaning, data integration and data extraction \cite{nargesian2019data}, and existing approaches cannot meet our trade-off between cost and quality.
They either demand domain specific training \cite{shin2015incremental}, which is more complicated in our scenario since it requires expert knowledge to annotate DBMS tuning knowledge, or they rely on strong assumptions that lack flexibility (e.g., focusing on specific document format) \cite{deng2022dom, lockard2019openceres, lockard2020zeroshotceres}. \textit{C2. Even with the prepared structured knowledge, we lack a way to integrate the knowledge into the optimization process.} The inherent design of optimization algorithms like Bayesian Optimization~\cite{snoek2012practical} and Reinforcement Learning~\cite{henderson2018deep} does not support the integration of external domain knowledge directly, necessitating extensive modifications to their standard workflows. For approaches that manually summarize static rules from domain knowledge, the resulting rules cannot capture the nuances of all workloads, and the updates of environments can make them out of date~\cite{10.5555/1287369.1287454, Kwan2002AutomaticCF}. 
To address the challenges above, we propose \textsc{GPTuner}, a manual-reading database tuning system that leverages domain knowledge automatically and extensively to enhance the optimization process.



\textit{Facing the challenge \textit{C1}, we develop a Large Language Model (LLM)-based pipeline to collect and refine heterogeneous knowledge, and propose a prompt ensemble algorithm to unify a structured view of the refined knowledge.} \color{black}In light of the brittle nature of LLM and its inevitable hallucination problem (see Section \ref{sec: motivation} for more details), we design an LLM-based pipeline incorporating two error correction mechanisms (i.e., step 2 is a filter of step 1 and step 4 is both an evaluator and a rewriter of step 3). \color{black} First, we prepare heterogeneous tuning knowledge from various resources (data ingestion). Second, we filter out noisy knowledge (data cleaning). Third, we summarize the multi-source knowledge by handling the possible conflict in a priority way (data integration). Fourth, we ensure the summarization result is factually consistent with source contents (data correction). Finally, we develop a prompt ensemble algorithm to construct a structured view of the knowledge (data extraction).


\textit{Regarding the challenge \textit{C2}, we use the structured knowledge to (1) design a workload-aware and training-free knob selection strategy, (2) develop a search space optimization technique considering the value range of each knob, and (3) propose a novel knowledge-based optimization framework to explore the optimized space.} Before the tuning process, we optimize the search space in terms of the space dimensionality and the values in each dimension. For space dimensionality, we leverage the text analysis ability of LLM to simulate the knob selection process of DBAs, considering the characteristics of DBMS, workload, specific query and knob dependencies. For values in each dimension, we discard meaningless regions, highlight promising space and take special situations into consideration. Next, we develop a novel Coarse-to-Fine Bayesian Optimization Framework. At first, since it is non-trivial to reduce the size of search space while still retaining the potential for optimal results \cite{feurer2019hyperparameter}, we seek help from domain knowledge to carefully design two search spaces of different granularity. Next, the two spaces are explored sequentially by BO from coarse granularity to fine granularity, with a bootstrap technique to serve as a bridging mechanism. This framework balances between the efficiency of coarse-grained search and the thoroughness of fine-grained search. 




We are aware of only one work DB-BERT \cite{10.1145/3514221.3517843} that utilizes a pre-trained language model to read manuals and uses the mined hints to guide a reinforcement learning algorithm. It exhibits rapid convergence as it benefits from the information gained via text analysis. However, it only yields sub-optimal performance since it utilizes the domain knowledge narrowly and suffers from the inadequate exploration of search space. \system tackles these limitations and thus achieves faster convergence and better performance improvement. As shown in Figure \ref{fig: intro-compare}, \system significantly outperforms DB-BERT on two representative benchmarks (TPC-H and TPC-C) with different optimization objectives (latency and throughput). 
It is important to note that \system is a distinct approach compared to DB-BERT, and its efficacy does not stem from simply substituting the language model. We verify this by replacing the BERT model in DB-BERT with GPT-4 and compare it with \system again. More details are provided in Section \ref{subsec: nlp-tuning}.

We compare \system against DB-BERT and other state-of-the-art approaches that do not use text as input. We consider different benchmarks (TPC-C and TPC-H), metrics (throughput and latency), and DBMS (PostgreSQL and MySQL). \system identifies better configurations in \textbf{16x} less time on average, achieving up to \textbf{30\%} performance improvement over the \textbf{best-performing} alternative. \color{black}In addition to performance comparison, we manually prepare and open-source two datasets\footnote{\color{black}\url{https://drive.google.com/file/d/1Ss6EL-B3lhKkwVNBW5vPu-JQ-IeldaUJ}\color{black}}, and conduct experiments to evaluate \textsc{GPTuner}'s robustness and scalability. Moreover, we discuss and present the overheads introduced in \system using different language models. \color{black} In summary, we make the following contributions:







\begin{itemize}
    \item We propose {\sc GPTuner}, a novel manual-reading database tuning system that leverages domain knowledge automatically and extensively to enhance the knob tuning process.
    \item We develop an LLM-based pipeline to collect and refine domain knowledge, and propose a prompt ensemble algorithm to unify a structured view of the refined knowledge.
    \item We design a workload-aware and training-free knob selection strategy, develop an optimization method for the value range of each knob, and propose a Coarse-to-Fine Bayesian Optimization framework to explore the optimized space.
    \item  We open-source the built domain knowledge for PostgreSQL and MySQL, \color{black}two datasets for the evaluation of LLM\color{black}, to enable the further study within the database community. 
    \item \color{black} We conduct experiments and costs analysis to evaluate the effectiveness, scalability and robustness of {\sc GPTuner}. \color{black}
\end{itemize}

\section{Background and Related Work}\label{sec: back-related-work}
\subsection{Database Knob Tuning} \label{subsec: knob-tuning}
DBMS expose tens to hundreds of configurable knobs to control their runtime behaviours \cite{10.14778/3450980.3450992}. The DBMS knob tuning problem is to select an appropriate value for each knob to optimize the DBMS performance (e.g., throughput or latency) on a certain workload (e.g., a workload is a set of SQL statements). Formally, given a set of configurable knobs $\theta_1,\dots,\theta_n$ along with their domains $\Theta_1,\dots,\Theta_n$, the configuration space is defined as $\boldsymbol{\Theta}=\Theta_1\times\dots\times\Theta_n$. We want to find a configuration $\boldsymbol{\theta}^{*}\in\boldsymbol{\Theta}$ to maximize DBMS performance $f$:
\begin{equation}
    \boldsymbol{\theta}^{*}= \argmax_{\theta \in \boldsymbol{\Theta}} f(\theta)
\end{equation}
Finding an optimal solution for this problem is challenging. Firstly, there are hundreds of knobs to tune in DBMS. Secondly, the knobs themselves vary widely -- some have continuous numerical values, while others are categorical, making the problem even more complicated since it is hard to model heterogeneous space. Three kinds of approaches are proposed: heuristic-based, Bayesian Optimization (BO)-based \cite{snoek2012practical} and Reinforcement Learning (RL)-based \cite{henderson2018deep}.

\noindent \labelitemi~ \textbf{Heuristic-based.}  Rule-based methods rely on manually created rules to explore the space in a predefined way \cite{10.5555/1287369.1287454, Kwan2002AutomaticCF}. Search-based methods explore the space according to several heuristics (e.g., avoid to revisit explored regions and explore nearby regions to improve the current optimum) \cite{10.1145/3127479.3128605, 10.1145/2628071.2628092}. 

\noindent \labelitemi~ \textbf{BO-based.} BO-based methods \cite{10.14778/1687627.1687767, 10.1145/3035918.3064029, 10.1145/3448016.3457291, 10.14778/3457390.3457404, 10.1145/3514221.3526176, 10.1145/3318464.3380591} follow the generic BO framework to search for an optimal configuration: (1) fitting a probabilistic \textit{surrogate model} to map the relation between knob configuration and DBMS performance, (2) selecting the next configuration that maximizes \textit{acquisition function}. 


\noindent \labelitemi~ \textbf{RL-based.} RL-based methods explore search space with a trial-and-error strategy. The essence is to balance between exploring unexplored space and exploiting known regions, which is achieved by the interactions between an agent (e.g., neural network) and the target environment (e.g., DBMS) \cite{10.1145/3299869.3300085, 10.14778/3352063.3352129, 10.1145/3514221.3517882, wang2021udo, 10.1145/3464389}.

\noindent \textbf{Performance Comparison.} According to \cite{10.14778/3538598.3538604}, a BO-based method, Sequential Model-based Algorithm Configuration (SMAC) \cite{lindauer-jmlr22a} yields the best performance since it is efficient in modeling the heterogeneous search space. \textit{We view it as the current state-of-the-art that does not take text as input and aim to improve it further.} 





\begin{figure*}[t]
  \centering
  \includegraphics[scale=0.146]{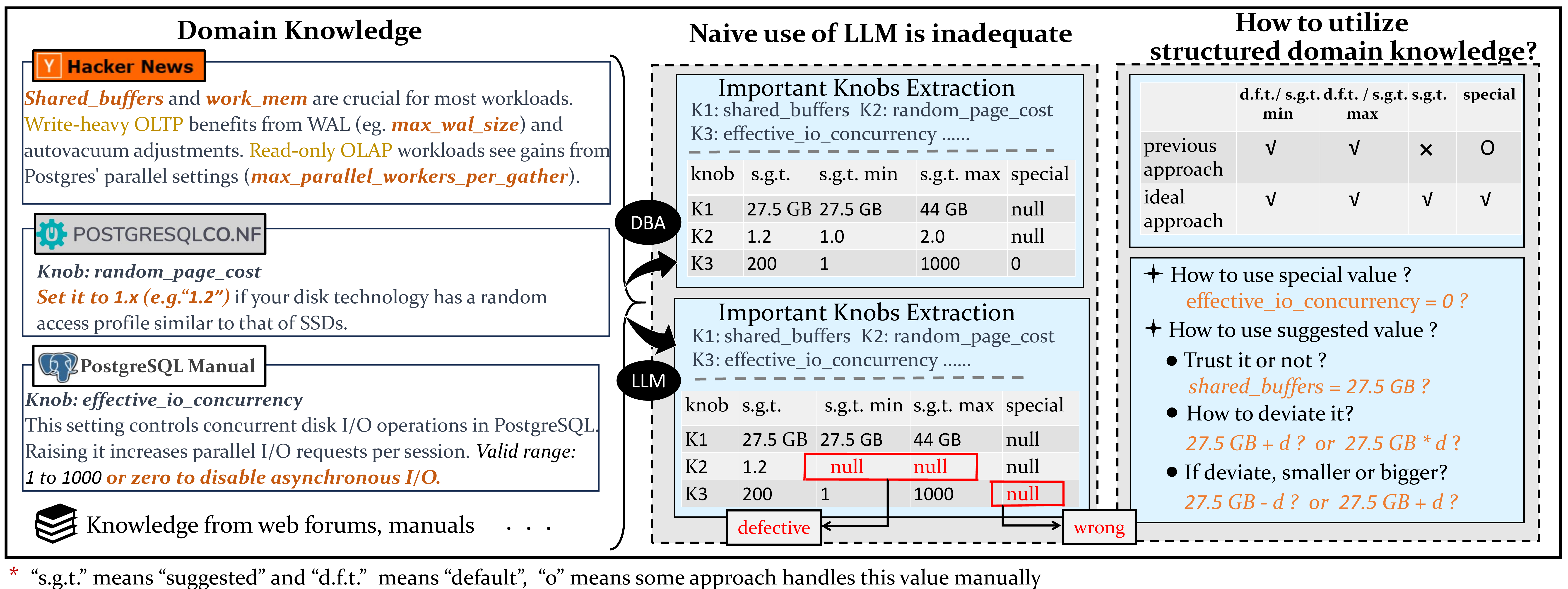}
   
  \caption{Motivating Example}
  \label{fig: motivation-example}  
   
\end{figure*}

\subsection{Language Models} \label{subsec: LLM}

The field of Natural Language Processing (NLP) has seen great progress with the advent of LLM. 
Notably, ChatGPT~\cite{luo2023chatgpt} and GPT-4~\cite{openai2023gpt4} have demonstrated great prowess across a wide range of tasks. The impact of LLM extends beyond the scope of NLP to the database community, experiencing a surge in the adoption of LLM to enhance DBMS, including data profiling~\cite{zhang2023schema, trummer2021can, narayan2022foundation}, code generation~\cite{trummer2023demonstrating, yu2019spider, cheng2023binding, scholak2021picard}, workload analysis~\cite{jain2018query2vec, wang2023real, preqr} and table-based question answering~\cite{ye2023large, zhang2023reactable, jiang2022omnitab}. When using LLM, there are three typical choices: (1) training a model from scratch, (2) fine-tuning an existing model, and (3) using a pre-trained model without parameter modifications. The first two options require a relatively large amount of resources, including hardware resources and training data. Given the impressive in-context learning ability of GPT-4, we have chosen the third option and use GPT-4 as our default model unless noted otherwise.

\begin{table}[H]
\centering
\caption{Comparison between DB-BERT and {\sc \bfseries GPTuner}}
 
\small
\setlength{\tabcolsep}{3pt}
\renewcommand{\arraystretch}{0.9}
\begin{tabularx}{\columnwidth}{lXX}
\toprule[1.2pt]
\textbf{Criterion} & \textbf{DB-BERT} & {\sc \bfseries GPTuner} \\
\specialrule{1.2pt}{\abovetopsep}{\belowrulesep}
Language Model & BERT & GPT-4 \\
\midrule
Workload-Aware Knob Selection & no & yes \\
\midrule
Fine-Tuning & yes & no \\
\midrule
Filter Noise & no & yes \\
\midrule
Space Type & Discrete & Heterogeneous \\
\midrule
Optimization Algorithm & Reinforcement Learning & Coarse-to-Fine BO \\
\midrule
Considered Knowledge & Suggested Value & \begin{minipage}[t]{\linewidth}
Suggested Value \\
Bound Constraint \\
Special Cases 
\end{minipage} \\
\bottomrule[1.2pt]
\end{tabularx}
\label{table: compare-dbbert-gptuner}
 
\end{table}

\subsection{Enhancing DBMS Knob Tuning with LMs} \label{subsec: nlp-tuning}

With the vast corpus of tuning guidance available on the internet or provided by DBMS vendors, language models can be harnessed to ``read the manual'' and provide structured tuning hints to enhance the knob tuning approaches. As the state-of-the-art LM-enhanced tuning approach, DB-BERT~\cite{10.1145/3514221.3517843}
models the tuning process as a series of ``multiple choice question answering problem'', and uses Reinforcement Learning to fine-tune a pre-trained BERT model to answer these problems. 
While DB-BERT converges fast as it benefits from the text analysis, it only yields sub-optimal performance. We summarize the main differences between \system and DB-BERT in Table \ref{table: compare-dbbert-gptuner}. \color{black}Please refer to our technical report \cite{technical_report} for more details.\color{black}

\section{Motivation}\label{sec: motivation}


\noindent \textbf{\color{black}M1: \color{black} ML-based methods still incur significant tuning costs.} State-of-the-art ML-based tuning methods require hundreds to thousands of iterations to reach a good DBMS knob configuration. Such high tuning expenses stem from the inefficiency of runtime feedback-based optimization algorithms. Specifically, the feedback information is limited (i.e., a few benchmark runs cannot provide a complete picture of the DBMS performance under all conditions) and unstable (i.e., DBMS performance is not guaranteed to improve after each step of knob tuning). Given such weak feedback, it takes a substantial number of observations for ML models to have sufficient confidence in predictions, especially when the  space is complex. 

\noindent \textbf{\color{black}M2: \color{black} Domain knowledge helps, but not well-exploited.} 
Extensive tuning knowledge has continually accumulated in the form of natural language and such knowledge is invaluable. As shown in Figure \ref{fig: motivation-example} (left part), we can identify which knobs are worth tuning and gain insights of the typical value settings for knobs (e.g., suggested values to try, range constraint and special values). However, such wisdom seems exclusive to DBAs. There are approaches that utilize the static rules summarized by DBAs to tune DBMS. Unfortunately, these rules cannot capture the nuances of all workloads, and the updates of environments can make these rules out of date. 

%
\begin{figure*}[t]
  \centering
  \includegraphics[scale=0.45]{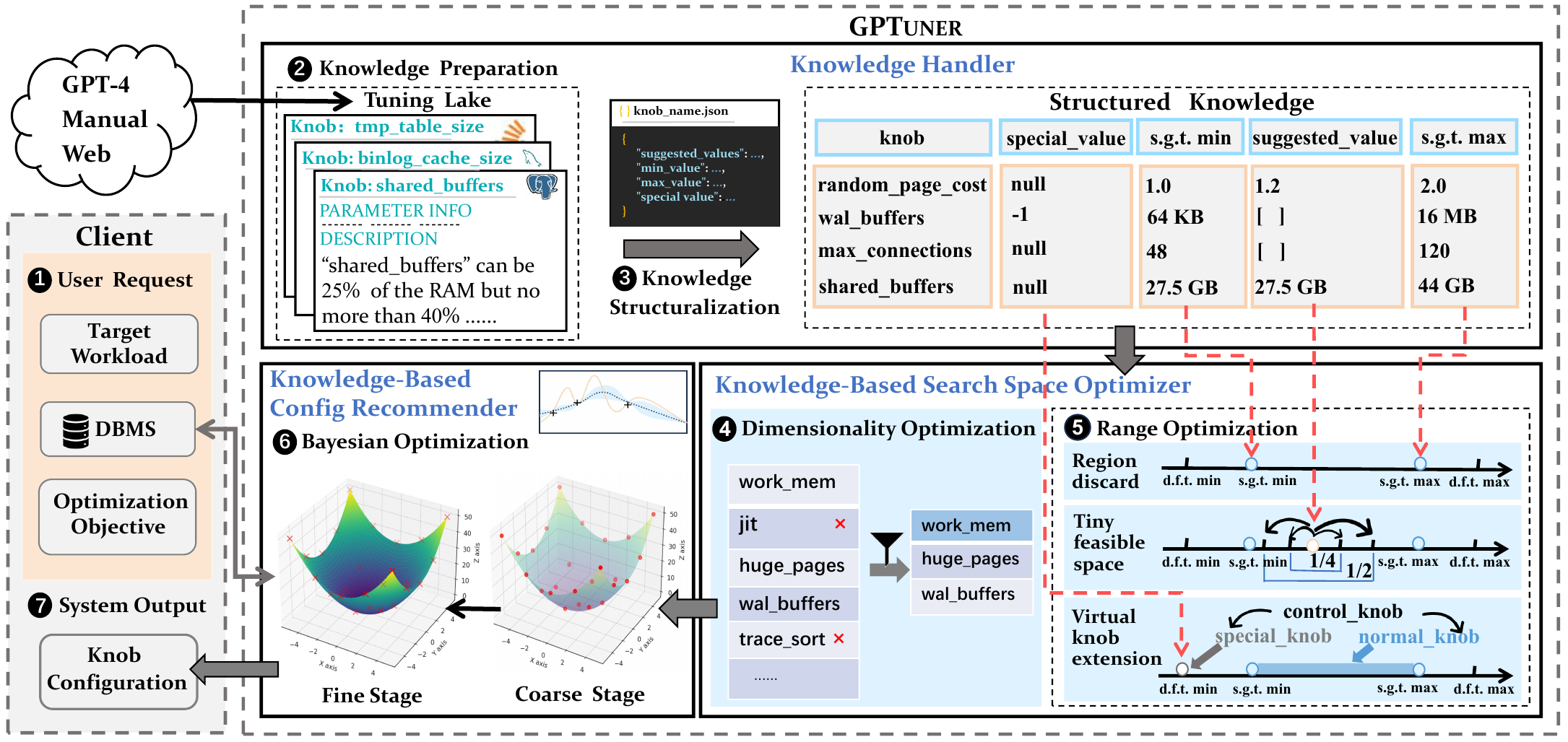}
   
  \caption{System Overview of {\sc \bfseries GPTuner} }
  \label{fig: gptuner-system}
   
\end{figure*}                                               
\noindent \textbf{\color{black}M3: \color{black} LLM is a notable step forward, but not adequate yet.} While it is acknowledged that domain knowledge is useful, such wisdom is considered inaccessible to machines due to the barriers in natural language understanding. Recently, the advent of LLM makes it possible to leverage such knowledge since we can utilize LLM to transfer the knowledge into a  machine-readable format (e.g., structured data like JSON and relational table).
However, this process is challenging. First, since domain knowledge typically comes in the form of DBMS documents and discussions from forums, it involves a complex and lengthy workflow to process such heterogeneous and noisy knowledge: data ingestion, data cleaning, data integration and data extraction. Second, the brittle nature of LLM (i.e. small modifications to the prompt can cause large variations in the outputs) and the hallucination problem of LLM (i.e., LLM generates answers that seem correct but are factually false) make it even more pronounced. For instance, as shown in Figure \ref{fig: motivation-example} (middle part), LLM can generate contents that are defective and even wrong. 



\noindent \textbf{\color{black}M4: The lack of a knowledge-aware optimization framework.} \color{black} To the best of our knowledge, no DBMS tuning framework considers structured knowledge. Algorithms like BO and RL do not support the integration of external knowledge directly, necessitating extensive modification to their standard workflows. The only information that can be used is the range constraint. However, there is much more useful information available (e.g., suggested values and special values). We present some simple ideas to utilize these values in Figure \ref{fig: motivation-example} (right part). Unfortunately, such naive ideas cannot serve as an effective solution, especially in the context of knob tuning where the problem is proven to be NP-hard \cite{10.1145/1005686.1005739}. 


\color{black}

\noindent \textbf{Outline.} To address the high tuning costs incurred in ML-based approaches (M1), we propose leveraging the invaluable but not well-exploited domain knowledge to enhance their efficiency (M2). However, this process is non-trivial and challenging, due to the limitations of LLM (M3) and the lack of a knowledge-aware optimization framework (M4). Therefore, we make the following technical contributions. To handle M3, we develop (1) an LLM-based pipeline with two error correction mechanisms in Section \ref{subsec: knowledge-extraction}, (2) a prompt ensemble algorithm in Section \ref{subsec: knowledge-transform}. To address M4, we develop (3) an LLM-based knob selection strategy in Section \ref{subsec: knob-select}, (4) three range optimization techniques in Section \ref{subsec: space-reduction}, and (5) a Coarse-to-Fine Bayesian Optimization Framework in Section \ref{sec: space-explorer}. Finally, we use an ablation study to reveal the benefit of each module in Section \ref{subsec: ablation}.

\color{black}

\section{Overview of GPTuner} \label{sec: GPTuner-design}

\noindent \textit{\underline{Workflow.}} \system is a manual-reading database tuning system to suggest satisfactory knob configurations with reduced tuning costs. Figure \ref{fig: gptuner-system} presents the tuning workflow that involves seven steps. \ding{182} User provides the DBMS to be tuned (e.g., PostgreSQL or MySQL), the target workload, and the optimization objective (e.g., latency or throughput). \ding{183} \system collects and refines the heterogeneous knowledge from different sources (e.g., GPT-4, DBMS manuals and web forums) to construct \textit{Tuning Lake}, a collection of DBMS tuning knowledge.
\ding{184} \system unifies the refined tuning knowledge from \textit{Tuning Lake} into a structured view accessible to machines (e.g., JSON). 
\ding{185} \system reduces the search space dimensionality by selecting important knobs to tune (i.e., fewer knobs to tune means fewer dimensions). 
\ding{186} \system optimizes the search space in terms of the value range for each knob based on structured knowledge. \ding{187} \system explores the optimized space via a novel Coarse-to-Fine Bayesian Optimization framework,
and finally \ding{188} identifies satisfactory knob configurations within resource limits (e.g., the maximum optimization time or iterations specified by users). 




\noindent \textit{\underline{Components.}} \system consists of three components: \textit{Knowledge Handler}, \textit{Knowledge-Based Search Space Optimizer} and \textit{Knowledge-Based Configuration Recommender} and they work as follows:

\noindent \textbf{Knowledge Handler.} \textit{Knowledge Handler} uses a DBMS tuning knowledge-oriented data pipeline to unify a structured view of the heterogeneous domain knowledge. At first, 
we propose an LLM-based pipeline to balance between cost and quality in Section \ref{subsec: knowledge-extraction}.
Next, we propose an \textit{LLM-based Prompt Ensemble Algorithm} to transfer the refined knowledge into a structured format such that it can be utilized by machines in Section \ref{subsec: knowledge-transform}.


\noindent \textbf{Knowledge-Based Search Space Optimizer.} The optimizer optimizes the search space from two aspects. Firstly, it reduces the size of the search space in terms of dimensionality. 
Specifically, we propose a workload-aware and training-free approach to select important knobs in Section \ref{subsec: knob-select}. Secondly, we optimize the search space by focusing on the value range of each knob. 
We propose \textit{Region Discard}, \textit{Tiny Feasible Space} and \textit{Virtual Knob Extension} methods to discard meaningless regions, highlight promising space and handle special situations, respectively (Section \ref{subsec: space-reduction}). 
    
\noindent \textbf{Knowledge-Based Configuration Recommender.} The recommender uses a novel Coarse-to-Fine BO Framework to compute optimal configurations. 
In the first stage, BO only explores a discrete subspace of the whole heterogeneous space. This subspace is small in size but promising to contain good configurations since we generate it based on the reliable domain knowledge. In the next stage, in order to avoid the overlooking problem of coarse-grained search (i.e., it is inevitable to lose some useful configurations for any space reduction technique), BO explores the heterogeneous space thoroughly with the optimizations in Section \ref{sec: space-explorer}. 
After the two stages, the recommender outputs the best-performing knob configurations found within the budget limits specified by users.

\section{Knowledge Handler} \label{sec: skill-library-adinistrator}


\subsection{Knowledge Preparation} \label{subsec: knowledge-extraction}

The Knowledge Preparation task is to collect tuning knowledge from various resources (data ingestion), filter out noisy contents (data cleaning), summarize the legal parts (data integration) and make sure the summarization is factual consistent with source contents (data correction). \color{black}We provide a running example in our technical report \cite{technical_report} and a demonstration \cite{gptunerdemo} for ease of understanding. \color{black} The ultimate output is a \textit{Tuning Lake} defined as follows:

\begin{definition}[Tuning Lake]
\label{definition: tuning-lake}
\textit{Tuning Lake} $\mathcal{L}=\{d_1, d_2,\dots d_n\}$ is a set of $n$ texts, where $n$ is the number of configurable knobs and $d_i$ is the natural language tuning knowledge for $i$-th knob. For example, ``set \texttt{shared\_buffers} to 25\% of the RAM'' (denoted as $d_i$) is the tuning knowledge for knob  ``\texttt{shared\_buffers}'' (the $i$-th knob). 
\end{definition}







    




\noindent \textit{Step 1: Extracting knowledge from LLM.} Except for the common tuning knowledge sources (e.g., manuals and web contents), we propose utilizing LLM as a knowledge source as well. Since GPT is trained on a vast corpus related to database \cite{ameryahia2023large}, GPT itself is an informative manual and allows us to retrieve the knowledge through prompt. In practice, we surprisingly find that GPT can give reasonable suggestions that are not included in the manuals. Such suggestions come from web contents summarized by DBAs and were used as training data for GPT. Since it is impossible to provide all web contents to any system and GPT already knows much of it, it is reasonable to use GPT as a complementary source of knowledge. In case that GPT gives nonsense suggestions, we utilize LLM to handle such abnormal situation in the next step.


%


\vspace{0.3em}

\noindent \textit{Step 2: Filtering noisy knowledge.} The tuning knowledge comes from various sources and its quality cannot be guaranteed. Thus we filter out noisy knowledge by modeling this process as a ``binary classification problem'' and utilize LLM to solve it. We provide LLM with the candidate tuning knowledge for a knob and an official system view for this knob (e.g., \textit{pg\_settings} from PostgreSQL and \textit{information\_schema} from MySQL). Moreover, we give a few examples in the prompt to utilize the in-context learning ability of LLM \cite{dai-etal-2023-gpt}. LLM evaluates whether the tuning knowledge conflicts with the system view and we discard any knowledge that does conflict.




%


\vspace{0.3em}

\noindent \textit{Step 3: Summarizing knowledge from various resources.} There can be multiple tuning knowledge for a knob. While such tuning guidance obeys the official system view (Step 2), they could conflict with each other (e.g., different manuals provide distinct recommended values for the same knob). We handle this by manually setting priority for each information source based on its reliability. For example, official manuals are authoritative and thus have the highest priority, while LLM has the lowest priority due to its hallucination problem. We summarize the non-contradictory guidance and delete the content with low priority for the contradictory parts. 



%

\vspace{0.3em}

\noindent \textit{Step 4: Checking factual inconsistency.} In the last step, the summary task is completed by LLM and the generated summaries may be factually inconsistent (i.e., the summary includes information that does not appear in the source contents or even contradicts it \cite{factuality}). Since GPT outperforms previous methods as a factual inconsistency evaluator \cite{luo2023chatgpt}, we utilize GPT to check this inconsistency. 
For each knob, the summarization and the source contents are included in prompt for GPT to determine whether there is an inconsistency. If an inconsistency is detected, GPT is prompted to recreate the summarization. This newly generated summary, along with its source contents, are once again provided to GPT. This process is repeated until GPT identifies no errors, producing the final summarization.




%

\color{black}
\noindent \textbf{Robustness Study.} It is important to note that the hallucination problem still remains a challenge in the NLP field, and thus we propose two error correction mechanisms (i.e., step 2 and 4 are designed to correct step 1 and 3, respectively) to minimize such impact as much as possible. To quantify their reliability, we manually prepare and open-source two datasets, and conduct experiments to measure the effectiveness of the two mechanisms. Moreover, we study the effect of domain knowledge quality (e.g., outdated and incorrect knowledge), and the LLM ability on \textsc{GPTuner}'s tuning performance. More details are provided in Section \ref{subsec: robust}.

\color{black}

\subsection{Knowledge Transformation} \label{subsec: knowledge-transform}

   
   


In this section, we build a structured view (namely \textit{Structured Knowledge}) of the \textit{Tuning Lake} such that it can be utilized by ML. Formally,
\begin{definition}[Structured Knowledge]
\label{definition: structured-knowledge}
\textit{Structured Knowledge} $\mathcal{S}$ maintains a structured view $s_i$ for each tuning knowledge $d_i$ from Tuning Lake $\mathcal{L}$, where $s_i$ is defined by a set of attributes $A=\{a_1, a_2, \dots, a_n\}$ (e.g., $a_1=suggested\_values$) with corresponding values $V=\{v_1,v_2,\dots,v_n\}$ (e.g., $v_1=4~GB$).
\end{definition}






\noindent \labelitemi~ \textbf{Determining the Attributes.} In the context of DBMS knob tuning, we primarily consider four types of attributes: \textit{suggested\_values}, \textit{min\_value}, \textit{max\_value} and \textit{special\_value}. Firstly, the most typical hint to consider is the recommended values for a knob since they performed well in the past practice and can serve as good starting points for new scenario. Secondly, we take into account the minimum and maximum values suggested in the tuning knowledge. This is because the default value ranges provided by DBMS vendors are excessively broad, complicating the optimization process (e.g., there will be a large number of values whose effect on the DBMS performance is unknown and this means ML models require much more iterations to converge to good configurations) and introducing the risk of system crashes (e.g., set memory-related knobs to a value higher than the available RAM can crash the DBMS). Finally, there are knobs with ``special values'' and these values are hard to be modeled by ML methods since they lead to distinct behaviours of DBMS. Thus we use \textit{special\_value} to handle such special situations. The techniques to utilize these values are detailed in Section \ref{subsec: space-reduction}.



\vspace{0.5em}

\noindent \labelitemi~ \textbf{Determining the Attribute Values.} Extracting specific knob values for certain attributes from given texts is challenging due to the brittle nature of LLM (i.e., small modifications to the prompt can cause large variations in the LLM outputs \cite{li2022advance}). Since it is useful to acquire a more reliable result by aggregating multiple imperfect but effective results \cite{arora2022ask, sagi2018ensemble}, we develop a \textit{Prompt Ensemble Algorithm} to determine the attribute values effectively.
Specifically, it involves three steps: modeling the extracting task as a Natural Language Problem such that it can be answered by LLM, varying the prompts to prepare multiple results, and aggregating these results to produce the final result:

Step 1: we model the transformation task as a series of information extraction problems.
At first, we decompose the transformation task into two subtasks of extracting (1) \textit{suggested\_values}, \textit{min\_value}, \textit{max\_value} and (2) \textit{special\_value}, respectively. We tackle \textit{special\_value} separately because \textit{special\_value} has its own context, which will be discussed in Section \ref{subsec: space-reduction}. Next, we prepare the prompt for each subtask. The $\{target\_values\}$ is a placeholder to be determined by task type (e.g., it is replaced with the definition of special values if we want to extract \textit{special\_value} from the $\{knowledge\}$). We also include examples in the prompt to utilize the in-context learning of LLM.



%

Step 2: we vary the prompts by changing the examples provided for few-shots learning. We manually prepare $K$ examples and store them in an \textit{Example Pool}. Then we randomly sample $n$ examples for each prompt ($n \leq K$). 
\color{black} Specifically, we manually prepare 10 examples ($K=10$) since 10 examples are empirically good enough to meet the diversity requirement of ensemble algorithms \cite{10.1145/3551349.3559555}. More examples can be added if domain experts are available. Regarding the $n$, more sampled examples might provide better results. However, this leads to a longer prompt with more tokens to be processed, subsequently resulting in longer processing time and higher monetary costs. Moreover, since too many examples might lead to long-context errors, we select $n=3$ as suggested in \cite{arora2022ask}. \color{black}

Step 3: we aggregate the results via a majority vote strategy. LLM generates a candidate JSON for each prompt and we aggregate the results by taking an \textit{Element-Wise Majority Vote} strategy, where \textit{Element} refers to our target attribute. For each attribute, we rank the values extracted based on occurrence frequency and choose the value of the highest frequency as the final result for that attribute. The resulting JSON is stored in the Structured Knowledge $\mathcal{S}$.

\section{Search Space Optimizer}\label{sec: space-optimizer}

\subsection{Dimensionality Optimization} \label{subsec: knob-select}
In this section, we identify knobs that have a significant impact on the DBMS performance and only tune these important knobs. 

\noindent \textbf{Motivation.} \color{black} While there are hundreds of knobs in DBMS (e.g. PostgreSQL v14.9 has 346 knobs), not all of them significantly impact DBMS performance, and it is infeasible to take all knobs into account due to the curse of dimensionality. Therefore, we only select important knobs to tune. \color{black}
Recent studies have shown that tuning a small number of knobs is enough to yield near-optimal performance while significantly reducing tuning costs \cite{10.14778/3450980.3450992, 10.5555/3488733.3488749}. Existing approaches rely on ML-based algorithms to select important knobs and this requires hundreds to thousands of evaluations on DBMS under different workloads and configurations \cite{10.5555/3488733.3488749, 10.14778/3538598.3538604, 10.1145/3035918.3064029}.
Differently, DBAs seek help from manuals to select which knobs are worth tuning and this yields better performance than ML-based algorithms \cite{10.14778/3450980.3450992}. However, it takes significant burden for DBAs to handle hundreds of knobs. 
Thus, we hope to remove the tedious burden by proposing a workload-aware and training-free approach via LLM. 
Specifically, we prompt LLM to simulate DBA's empirical judgement in real-world scenarios by considering the four factors:

\textit{(1) System-Level selects knobs based on the specific DBMS product.} After years of tuning practice, it is empirically known which knobs are important for a certain DBMS product. For instance, it is widely discussed that we can gain substantial performance improvement by tuning ``\texttt{shared\_buffers}'' and ``\texttt{max\_wal\_size}'' from PostgreSQL, ``\texttt{innodb\_buffer\_pool\_size}'' and ``\texttt{innodb\_log\_file\_size}'' from MySQL \cite{10.14778/3450980.3450992}. More importantly, we find such wisdom is included in the corpus of GPT-4 \cite{openai2023gpt4} and we can extract it by prompting GPT-4 to recommend knobs to tune based on the DBMS product.

\textit{(2) Workload-Level selects knobs based on the workload type.} The workload type is informative for selecting important knobs since different workload types have distinct requirements on DBMS resources, which are regulated by knobs. For example, one typical scenario for OLTP workload is I/O-intensive, where write operations compete for the limited disk I/O. Thus we should consider disk-related knobs like ``\texttt{effective\_io\_concurrency}''. For analytical queries from OLAP workload, we should tune planning-related and system resource-related knobs to handle their complex structure and resource-intensive nature \cite{10106050}. 

\textit{(3) Query-Level selects knobs based on the bottleneck of queries.} It is useful for DBAs to delve into the execution plan of the time-consuming and frequently executed queries, where an experienced DBA can diagnose performance bottlenecks and focus on related knobs. Equipped with the analysis ability of LLM \cite{zhou2023llm}, we can include the execution plan in the prompt such that LLM can choose the bottleneck-aware knobs. For example, if LLM detects ``Sequential Scan'' in the execution plan and the target table contains a large number of rows, LLM will recommend adjusting scan-related knobs like ``\texttt{random\_page\_cost}'' since it influences the PostgreSQL query planner's bias towards index scans or sequential scans.

\textit{(4) Knob-Level complements interdependent knobs to a given target knob set.} One important reason why ML techniques outperform DBA is that they can handle the dependencies between knobs \cite{10.14778/3450980.3450992}. However, if the given knob set contains the important knobs but excludes the interdependent knobs, such ability is wasted. Since many dependencies are explicitly mentioned in the manuals, we can leverage the text analysis ability of LLM to read the manuals and capture such knobs. For example, the official PostgreSQL document suggests ``Larger settings for `\texttt{shared\_buffers}' usually require a corresponding increase in `\texttt{checkpoint\_segments}' '' \cite{PostgreSQL2023doc}, indicating that we should consider the two knobs at the same time.

Based on the above four factors, we develop the \textbf{LLM-based Knob Selection} as Algorithm \ref{algo: knob-select}. It aspires to utilize LLM $\mathcal{F}$ to select important knobs $\mathcal{T}$ from the Knob Set $\mathcal{K}$ for a DBMS $\mathcal{D}$ under a specific workload $\mathcal{W}$. The algorithm starts by preparing a configurable knob set $\mathcal{C}$, which excludes knobs related to debugging, security and path-setting (Line 1). Next, it selects knobs from four different levels: System Level, Workload Level, Query Level and Knob Level. In the System Level Selection, knobs hugely influencing the DBMS performance are identified, producing the set $\mathcal{C}_s$ (Line 2). During the Workload Level Selection, we provide LLM with the type of the workload (e.g., OLTP or OLAP) and the optimization target (e.g., throughput or latency) to prepare knob set $\mathcal{C}_w$ (Line 3). The Query Level Selection delves into each query $q_i$ within the workload (Line 4-9). Specifically, the execution plan $\mathcal{E}_i$ of each query is retrieved, and LLM $\mathcal{F}$ extracts influential knobs set $\mathcal{C}_q$ by diagnosing the performance bottlenecks of each query and identifying knobs related to the bottleneck. Finally, we leverage LLM $\mathcal{F}$ to replenish interdependent knobs to the  union of $\mathcal{C}_s$, $\mathcal{C}_w$ and $\mathcal{C}_q$, resulting in the final Target Knob Set $\mathcal{T}$(Line 10).

\begin{algorithm}[t]
    \SetAlgoLined
    \caption{LLM-based Knob Selection}
    \label{algo: knob-select}

    \SetKwInput{Input}{Input}
    \SetKwInput{Output}{Output}
    \SetKwInput{Workload}{Workload Level Selection}
    \SetKwInput{Query}{Query Level Selection}
    \SetKwInput{System}{System Level Selection}
    \SetKwInput{Knob}{Knob Level Selection}
    
    \Input{Knob Set $\mathcal{K}$; LLM $\mathcal{F}$; DBMS $\mathcal{D}$; Workload $\mathcal{W}$; Tuning Lake $\mathcal{L}$.}
    \Output{Target Knob Set $\mathcal{T}$.}

    Configurable Knob Set $\mathcal{C} \gets FILTER(\mathcal{K})$\; \tcp{Filter out knobs that are related to debugging, security and path-setting}

    \System{}
    $\mathcal{C}_{s} \gets \mathcal{F}(\mathcal{C}, \mathcal{D})$\;

    \Workload{}
    $\mathcal{C}_{w} \gets \mathcal{F}(\mathcal{C}, \mathcal{W})$\;

    \Query{}
    $\mathcal{C}_q \gets \emptyset$\;
    \For{query $q_i$ ~ in ~$\mathcal{W}$}{
        $\mathcal{E}_i \gets EXECUTE(\mathcal{D},q_i)$\; \tcp{Get execution plan for query $q_i$ from $\mathcal{D}$}
        $\mathcal{C}_{q_{i}} \gets \mathcal{F}(\mathcal{C}, \mathcal{E})$\;
        $\mathcal{C}_q \gets$ $\mathcal{C}_q \cup \mathcal{C}_{q_{i}}$\;
    }

    \Knob{}
    Target Knob Set $\mathcal{T} \gets \mathcal{F}(\mathcal{L}, ~\mathcal{C}_{s} \cup \mathcal{C}_{w} \cup  \mathcal{C}_{q}$)\;

    \Return $\mathcal{T}$\;
\end{algorithm}

\subsection{Range Optimization} \label{subsec: space-reduction} 


In this section, given each knob's unique semantics and associated tuning knowledge, we optimize the value range for each knob. 


 
\noindent \textbf{Region Discard.} We utilize \textit{min\_value} and \textit{max\_value} to discard some regions for the following cases. (1) The regions are unlikely to result in promising performance. For knob ``\texttt{random\_page\_cost}'', the value range regulated by DBMS is $[0, 1.79 \times 10^{308}]$. 
Given the large value range, the algorithm is likely to sample very large values. This will lead to poor DBMS performance since it is suggested to set it to ``$1.x$'' if the disk has a random access profile similar to that of SSDs \cite{PostgreSQLCONF}. Equipped with this prior knowledge, we release the burden for optimizers to trial vast but meaningless space. (2) The regions could seize too many system resources. For resource-related knobs like \texttt{shared\_buffers}, a value close to the maximum system resource could be detrimental to other services that are running on the same machine. (3) The regions that can make the DBMS crash. For resource-related knobs, setting a value that exceeds the resource limits could prevent the DBMS from starting (e.g., we cannot set memory-related knobs to a value higher than the available RAM). Finally, the recommended range $[min\_value, max\_value]$ is much more narrow than the range provided by DBMS vendors.

\noindent \textbf{Tiny Feasible Space.} We use \textit{suggested\_values} from Structured Knowledge to define a discrete space for each knob. Such values are valuable since they performed well in the past and can serve as good starting points for new scenario. However, they may not be suitable for all cases, as the optimal knob setting depends on the specific environment, which can be diverse.
Instead of relying on these values only, we can apply a set of multiplicators for each suggested value of all numerical knobs. The intuition behind is to deviate the suggested value in different directions (smaller or bigger) with different extents. Formally, given a set of multiplicators $M=\{m_1, m_2, \dots, m_n\}$ and a suggested value $V$ for a knob $k$, the search space $\Omega$ for this knob is defined as:
\begin{equation}
    \Omega(k) = \{\alpha \cdot V | ~ \alpha \in M \}
\end{equation}
However, this heuristic approach ignores the value ranges of knobs and the multiplication results could be useless. For example, knob ``\texttt{checkpoint\_timeout}'' has a value range from 30 to 86400 with 90 as a recommended starting point \cite{enterprisedb}. Given the maximum factor used in DB-BERT (i.e., 4) \cite{10.1145/3514221.3517843}, the maximum value to try is $90 \times 4 = 360$ and this is much smaller than $86400$, which means a lot of promising values are ignored. Thus we address this limitation by considering the value range and calculating the multiplicators dynamically. For knob $k$, we denote its maximum (minimum) value as $U$, the multiplicator is calculated by the following formula:
\begin{equation}
    \alpha = 1 + ~ \frac{\beta}{V}(U-V),~\beta \in \{r_1, r_2, \dots, r_n ~|~ r_i \in [0, 1]\}
\end{equation}
where $\beta$ is a scaling coefficient with a value from 0 to 1, and the $n$ candidate values $r_i$ are predefined by users. The choice of $U$ determines the deviation direction (e.g., maximum makes value bigger while minimum makes it smaller) and $\beta$ controls the changing extends. Specifically, $V$ remains the same or is extended to the maximum (minimum) when $\beta$ is set to 0 and 1, respectively. For $\beta$ value between 0 and 1, suggested value $V$ moves proportionally to its maximum (minimum). In our experiments, $\beta\in\{0, 0.25, 0.5\}$. We conduct this deviation process for all numerical knobs and the resulting discrete space is our \textit{Tiny Feasible Space}, where \textit{Tiny} means the possible number of values for knobs is significantly reduced, and \textit{Feasible} indicates the chosen values are promising.


\noindent \textbf{Virtual Knob Extension.} This extension is to handle knobs that use special values to do something different from what the knobs normally do \cite{10.14778/3551793.3551844, 10.14778/3476311.3476411}. For example, knob ``\texttt{lock\_timeout}'', with a value range from 0 to 2147483647, controls the maximum allowed duration of any wait for a lock. When it is set to zero, the timeout function is disabled, making ``0'' a special value. However, optimizers may never trial this value (even though it could be optimal) since the likelihood of sampling it is extremely low \cite{10.14778/3551793.3551844} and the DBMS performance will be modeled to degrade as the knob value converges to zero \cite{10.14778/3476311.3476411}.
Thus, we develop a \textit{Virtual Knob Extension} technique as outlined in Algorithm \ref{algo: virtual-knob}. Firstly, we utilize \textit{Structured Knowledge} to select which knobs have the special values (Line 1). Such information is explicitly provided in official documents \cite{PostgreSQL2023doc} and thus can be understood and summarized by \textit{Knowledge Handler} (Section \ref{subsec: knowledge-extraction}). Secondly, we add ``virtual knobs'' (\textit{control\_knob}, \textit{normal\_knob} and \textit{special\_knob}) for these knobs (Line 3). \textit{control\_knob} is a knob with a value of zero or one to determine which value range will be used (Line 6). \textit{normal\_knob} and \textit{special\_knob} represent the normal and the special value range of the original knob, respectively. Based on the value of \textit{control\_knob}, only one of \textit{normal\_knob} and \textit{special\_knob} will be activated (Line 7-11). 
Above technique makes the special values of knobs to be considered by optimizers.

\begin{algorithm}[h]
    \SetAlgoLined
    \caption{Virtual Knob Extension}
    \label{algo: virtual-knob}

    \SetKwInput{Input}{Input}
    \SetKwInput{Output}{Output}
    \Input{Knob Set $\mathcal{K}$; Skill Library $\mathcal{S}$; Optimizer $\mathcal{O}$.}
   
    \SetKwInput{generate}{Virtual Knobs Generation}
    \SetKwInput{utilize}{Virtual Knobs Utilization}

    \generate{}
    Use $\mathcal{S}$ to identify special knobs $\mathcal{R} \subset \mathcal{K}$ \;
    \For{$r \in \mathcal{R}$}{
         Create virtual knobs $\{control\_knob, normal\_knob, special\_knob\}$ for $r$\;
    }
    
    \utilize{}
    \For{round = 1 to number of iterations}{
        $\mathcal{O}$ chooses a value from $\{0,1\}$ for $control\_knob$\;
        \uIf{\textit{control\_knob} is 0}{
            Activate \textit{normal\_knob} in $\mathcal{O}$\;
        }
        \Else{
            Activate \textit{special\_knob} in $\mathcal{O}$\;
        }
    }   
\end{algorithm}


%

\section{Configuration Recommender}\label{sec: space-explorer}



In this section, we discuss how to search the optimized space. 


\noindent \textbf{Basic Idea of Bayesian Optimization.} Bayesian Optimization is a sequential model-based algorithm~\cite{snoek2012practical}. 
It consists of two core components: the \textit{surrogate model} and the \textit{acquisition function}. A surrogate model takes a configuration as input and predicts the DBMS performance. The acquisition function evaluates the utility of candidate configurations (e.g., Probability of Improvement (PI) or Expected Improvement (EI)), and we choose the next configuration with the maximum utility. After the surrogate model is initialized, BO works iteratively as follows: (1) selecting the next configuration by maximizing acquisition function, (2) evaluating the configuration on DBMS and updating the surrogate model with the new observation. This process is repeated until it runs out of resources.


\noindent \textbf{Limitation of Bayesian Optimization.} The key to the success of BO-based methods is its surrogate model.
Recent works utilize random forest as the surrogate model to leverage its efficiency in modeling high-dimensional and heterogeneous search space, achieving the state-of-the-art performance \cite{10.14778/3538598.3538604, 10.1145/2487575.2487629}. 
However, the number of samples required to have sufficient confidence in predictions could still be significant \cite{10.1145/3318464.3380591, 10.14778/3551793.3551844}. 
Specifically, it requires hundreds of iterations to achieve satisfactory performance, which is resource-intensive and time-consuming. The key observation of this work is that such iteration cost could be significantly reduced if we integrate the domain knowledge into the optimization process, which motivates the following novel optimization framework. 

\begin{algorithm}[b]
    \SetAlgoLined
    \caption{Coarse-to-Fine BO Framework. \textcolor{blue}{Blue underlined text highlights differences to original BO algorithm}.}
    \label{algo: two-stage-BO-framework}

    \SetKwInput{Input}{Input}
    \SetKwInput{Output}{Output}
    \Input{DBMS $\mathcal{D}$; Surrogate Model $\mathcal{M}$; Acquisition Function $\mathcal{A}$; Workload $\mathcal{W}$; Structured Knowledge $\mathcal{S}$; Whole Space $\mathcal{P}$; Coarse Threshold $\mathcal{C}$; Initial Number $n$.}
    \Output{Knob Configuration $\mathcal{X}$.}

    \SetKwInput{Coarse}{Coarse Stage}
    \SetKwInput{Fine}{Fine Stage}

    {\color{blue} \underline{Generate Tiny Feasible Space $\mathcal{T}$ from $\mathcal{S}$}\;}

    Generate $n$ samples {\color{blue} \underline{$p_i \in \mathcal{T}$}} with space-filling design (LHS)\;

    Evaluate samples on $\mathcal{D}$ to get performance $y_i$\;

    Update $\mathcal{M}$ with observations $\{(p_i, y_i)\}$\;

    \Coarse{}
    \For{$i=1$ to $\mathcal{C}$}{
        $\vec{x_i} \gets \argmax_{\color{blue} \underline{\vec{x} \in \mathcal{T}}} \mathcal{A}(\vec{x})$ \;

        Evaluate $\vec{x_i}$ on $\mathcal{D}$ to get performance $y_i$\;
        
        Update $\mathcal{M}$ with $(\vec{x_i},~y_i)$\;
    }
    
    \Fine{}
    {\color{blue} \underline{Reuse the surrogate model $\mathcal{M}$ from Coarse Stage\;}}
    
    {\color{blue} \underline{Apply \textit{Region Discard} on $\mathcal{P}$ to get $\mathcal{P'}$\;}}
    
    {\color{blue} \underline{Apply \textit{Virtual Knob Extension} on $\mathcal{P'}$ to get $\mathcal{P''}$\;}}
    
    
    \While{not stopping condition}{
        $\vec{x_i} \gets \argmax_{\color{blue} \underline{\vec{x} \in \mathcal{P''}}} \mathcal{A}(\vec{x})$\;

        Evaluate $\vec{x_i}$ on $\mathcal{D}$ to get performance $y_i$\;
        Update $\mathcal{M}$ with $(\vec{x_i},~y_i)$\;
    }

    $\mathcal{X} \gets \argmax_{\vec{x_i}} y_i$\;

    \Return $\mathcal{X}$\;
\end{algorithm}

\noindent \textbf{Coarse-to-Fine Bayesian Optimization Framework.} The intuition behind is to take advantages of the efficiency of coarse-grained search (i.e., we aim to acquire non-optimal but effective solutions with low expenses) and the thoroughness of fine-grained search (i.e., we aim to find the possible optimal solutions by exploring the space thoroughly).
Algorithm \ref{algo: two-stage-BO-framework} describes the workflow and highlights the main differences from the basic BO algorithm. 


\textit{Coarse-grained Stage.} In the first stage, we only explore part of the whole space. It is a widely adopted approach to discretize the value ranges of parameters evenly for coarse-grained search. However, we will lose too many promising solutions  because such space reduction technique is inherently imprecise and non-adaptive. We seek help from knowledge to reduce the space while still retaining the potential for optimal results. Specifically, we explore the \textit{Tiny Feasible Space} (Line 1) defined in Section \ref{subsec: space-reduction}, which is small but reliable because it comes from manuals. Following previous works \cite{10.14778/1687627.1687767, 10.14778/3551793.3551844, 10.14778/3538598.3538604}, we initialize the surrogate model with ten samples ($n=10$) generated by Latin Hypercube Sampling (LHS) \cite{10.1145/167293.167637}, which distributes samples evenly across the whole space. Instead of sampling points from the whole space $\mathcal{P}$, we sample from the Tiny Feasible Space $\mathcal{T}$ (Line 2). After the samples are evaluated on $\mathcal{D}$ (Line 3) and the surrogate model is initialized (Line 4), we explore $\mathcal{T}$ with the BO algorithm for $\mathcal{C}$ iterations (Line 5-8). After the coarse-grained stage, we try out $n+\mathcal{C}$ configurations and this already yields non-optimal but promising results in practice, owing to the guidance of domain knowledge.

\textit{Fine-grained Stage.} Since it is inevitable to lose some important configurations for any space reduction technique, we explore the space thoroughly until we run out of resource limits or reach expected performance improvement (Line 13-17). However, this process could be exhausting, especially when there are hundreds of knobs to tune. Thus we make optimizations to enhance the search: (1) we bootstrap BO with the samples from the first stage (Line 10), (2) we narrow down space $\mathcal{P}$ with the \textit{Region Discard} technique (Line 11), and (3) we take into account the knobs with special values with the \textit{Virtual Knob Extension} technique (Line 12). \color{black} Please refer to our technical report \cite{technical_report} for a detailed discussion on our bootstrap. \color{black}

\section{Experimental Evaluation}\label{sec:experiments}

\subsection{Experimental Setup}





\noindent \textbf{Workloads.} Following the setup of DB-BERT \cite{10.1145/3514221.3517843}, we use TPC-H (OLAP type) with scale factor 1 and TPC-C (OLTP type) with factor 200 as our benchmarks. TPC-C uses ten terminals with unlimited arrival rate and the implementations are from BenchBase \cite{DifallahPCC13}. 

\noindent \textbf{Hardware.} We conduct our experiments using a cloud server with a 24-core Intel Xeon E5-2676 v3 CPU, 110 GB of RAM and a 931 GB WD SSD. We utilize a GeForce RTX 2080 Ti with 22 GB of memory. 


\noindent \textbf{Baselines.} \system is implemented with SMAC3 library  \cite{JMLR:v23:21-0888} and uses OpenAI completion API \cite{openai2023gpt4}. We compare it with state-of-the-art methods: DDPG++ \cite{10.14778/3450980.3450992} is a RL-based approach, GP is a Gaussian Process-based optimizer used in iTuned \cite{10.14778/1687627.1687767} and OtterTune \cite{10.1145/3035918.3064029}, SMAC \cite{10.14778/3538598.3538604} is the best-performing BO-based method with random forest as its surrogate, and DB-BERT is a database tuning tool \cite{10.1145/3514221.3517843} that uses BERT for text analysis to guide a RL algorithm.

\noindent \textbf{Tuning Settings.} We run experiments with PostgreSQL v14.9 and MySQL v8.0. Following previous works \cite{10.1145/3299869.3300085, 10.14778/3551793.3551844, 10.14778/3538598.3538604}, all algorithms tune the same 60 and 40 knobs of PostgreSQL and MySQL respectively.
We run three tuning sessions for each method, with each session consisting of 100 iterations and each iteration requires a stress test for the target workload. Aligning with the latest experimental evaluation \cite{10.14778/3538598.3538604}, we optimize throughput for OLTP workload and 95-th percentile latency for OLAP workload, reporting the median and quartiles of the best performances. \color{black} Please refer to our technical report for more details on the tuning settings \cite{technical_report}. \color{black}

\subsection{Performance Comparison}\label{subsec: performance-compare}

\begin{figure}[h]
  \centering
  \includegraphics[scale=0.25]{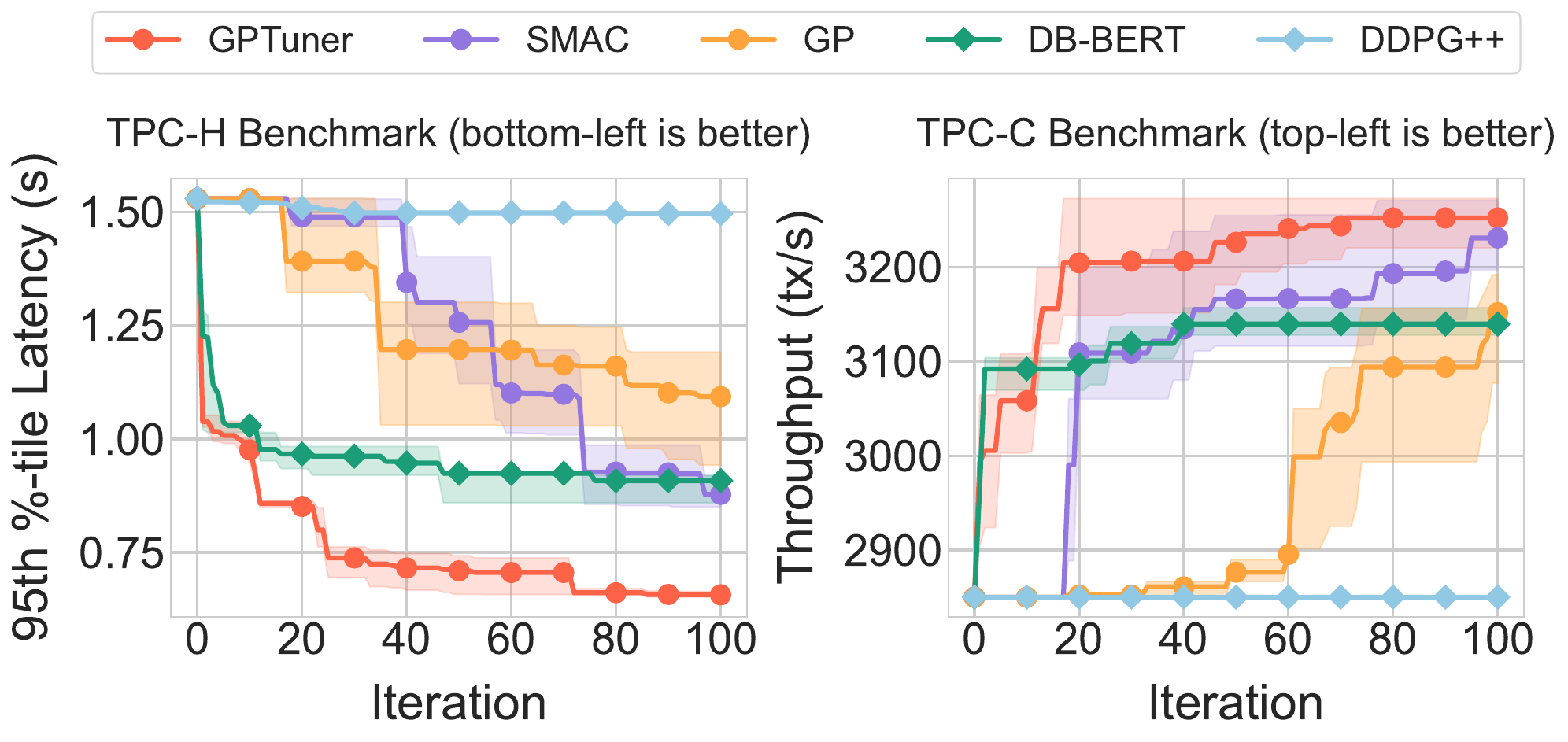}
   
  \caption{Best performance over iterations on PostgreSQL}
  \label{fig: postgres}  
\end{figure}

\noindent \textbf{Optimizing for PostgreSQL.} 
As shown in  Figure \ref{fig: postgres}, \system finds the best performing knob configuration with significantly less iterations in both benchmarks. For example, \system rapidly achieves significant performance improvement and reaches near-optimal latency (44.4\% less latency) with only 20 iterations in terms of TPC-H benchmark. 
Moreover, \system continues to achieve a 12.7\% reduction of latency after the 20-th iteration. While DB-BERT presents similar performance in the first 20 iterations (37.5\% less latency), it fails to further reduce the latency (only 3.4\% reduction). For methods that do not use domain knowledge, they converge much slower than \system (26x slower on average) and fail to find a configuration comparable to \textsc{GPTuner} (35.6\% worse on average).
The results on TPC-C are similar to that on TPC-H, where \system converges 12.6x faster and reaches the highest throughput. On average, \system takes 92.6\% less optimization time to achieve the best performance of the baselines. In the rest of the experiments, we focus on \textit{sample efficiency} only (see Section \ref{subsec: cost-analysis} for a discussion).

\begin{figure}[h]
  \centering
  \includegraphics[scale=0.25]{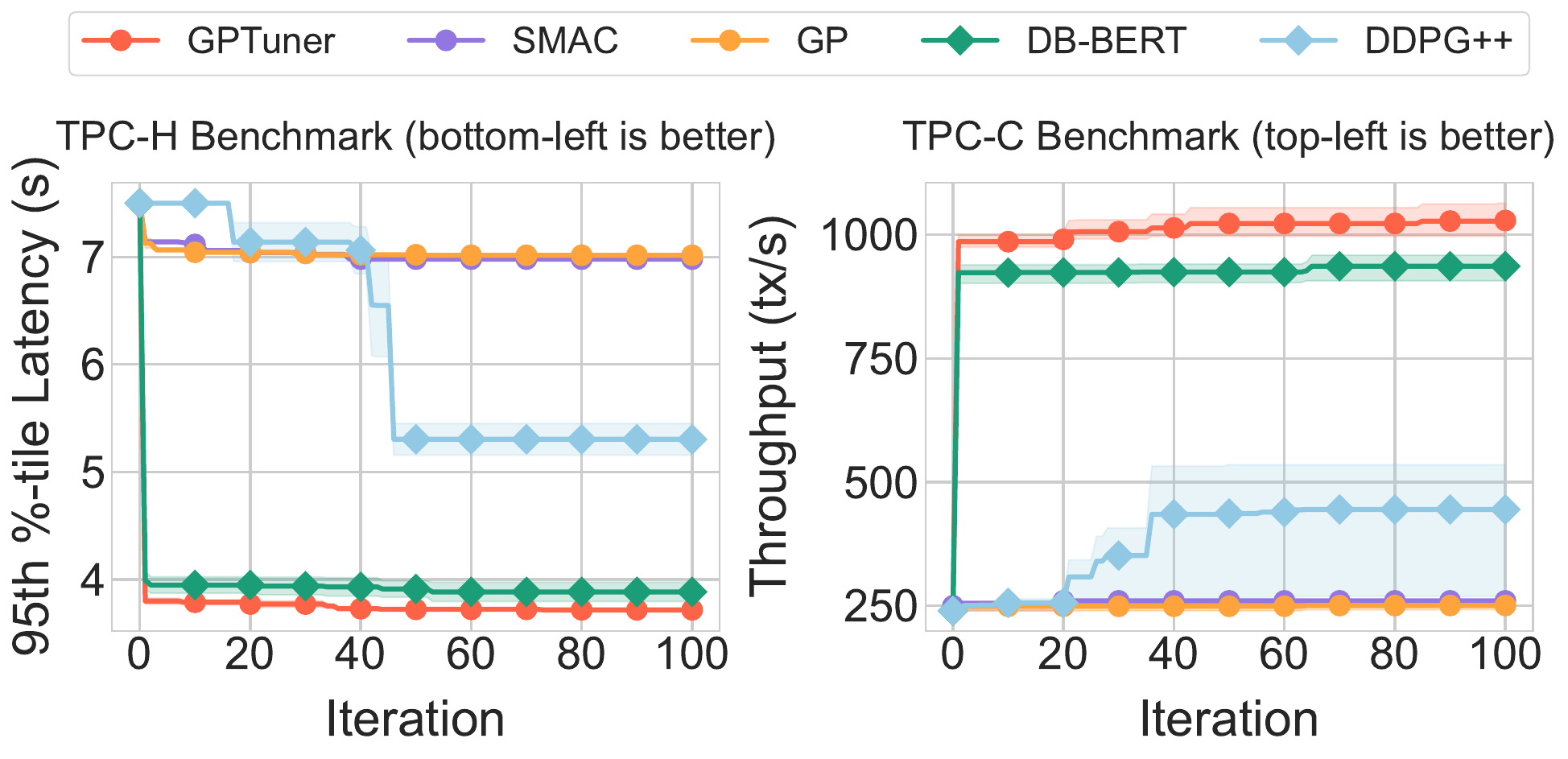}
   
  \caption{Best performance over iterations on MySQL}
  \label{fig: mysql}  
   
\end{figure}

\noindent \textbf{Optimizing for MySQL.}
The results on MySQL are presented in Figure \ref{fig: mysql}, which are similar to that on PostgreSQL. In terms of TPC-H, \system significantly reduces the latency by 48.5\% at the very beginning, surpassing the best performance achieved by all other baselines within 100 rounds. For TPC-C, \system achieves the highest throughput of 1029 tx/s over the default 240 tx/s, which is 10\% and 131.7\% better than DB-BERT and DDPG++, respectively. Notably, GP and SMAC fail to have considerable performance improvement on both benchmarks. This is because the default value ranges provided by DBMS vendors are excessively broad, making the optimizers struggle to explore the vast search space without the guidance of domain knowledge.

\captionsetup[subfloat]{font=small}

\noindent\begin{figure*}[!thb]

    \begin{minipage}{0.76\textwidth}
        \begin{figure}[H]
            \centering

        \includegraphics[scale=0.25]{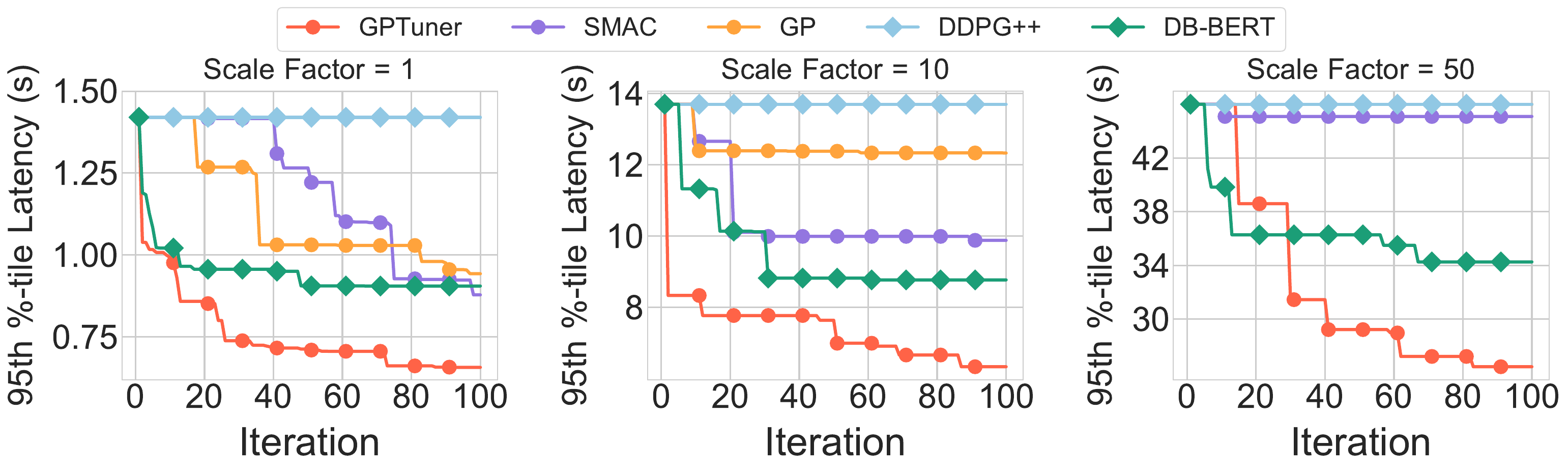}

            \caption{\color{black}Effect of Database Size on Tuning Performance (bottom-left is better)\color{black}}
            \label{fig: effect-data-size}
        \end{figure}
    \end{minipage}%
    \hfill
    \begin{minipage}{0.24\textwidth}
        \begin{figure}[H]
            \centering
            \includegraphics[scale=0.25]{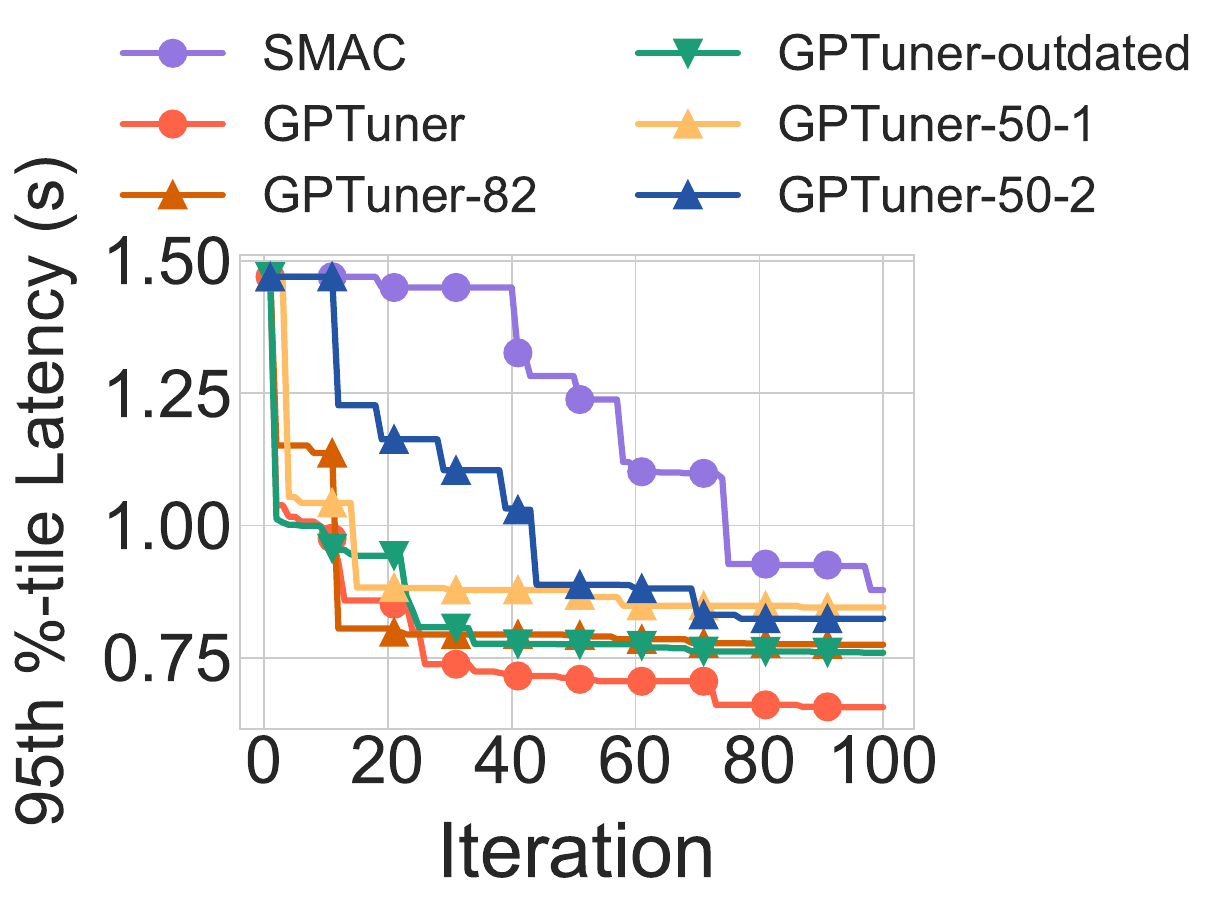}
             
            \caption{\color{black}Effect of Knowledge Quality (bottom-left is better)\color{black}}
            \label{fig: knowledge-quality}
        \end{figure}
    \end{minipage}
 
\end{figure*}

\noindent\begin{figure*}[!thb]
    \begin{minipage}{0.76\textwidth}
        \begin{figure}[H]
            \centering

        \includegraphics[scale=0.25]{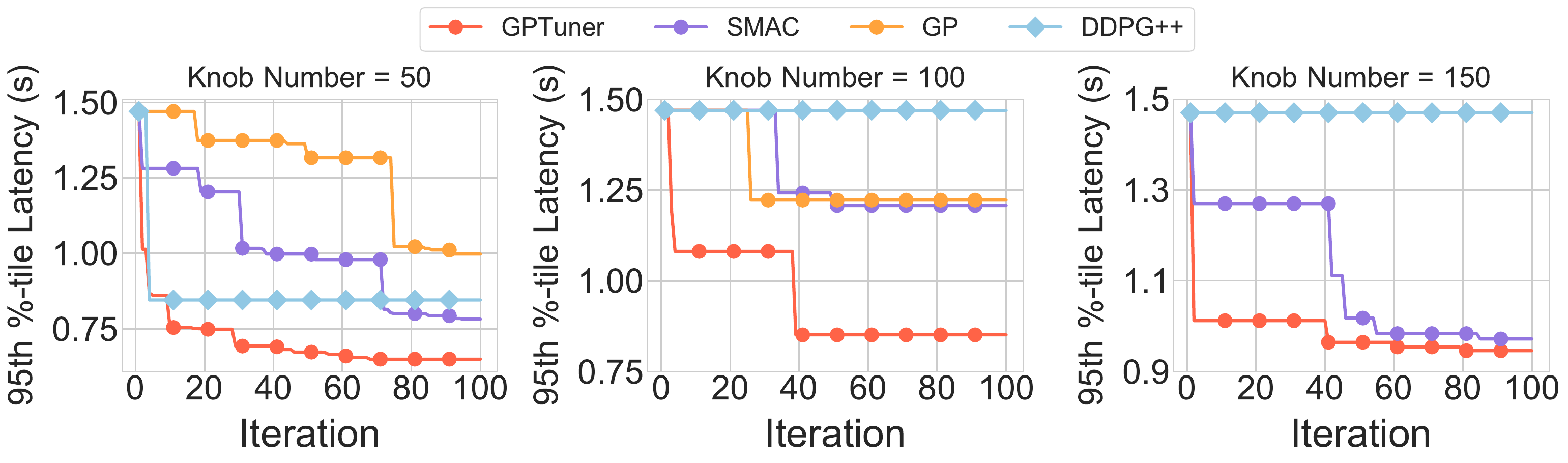}

            \caption{\color{black}Effect of Space Dimensionality on Tuning Performance (bottom-left is better)\color{black}}
            \label{fig: effect-space-dimension}
        \end{figure}
    \end{minipage}%
    \hfill
    \begin{minipage}{0.24\textwidth}
        \begin{figure}[H]
            \centering

            \includegraphics[scale=0.25]{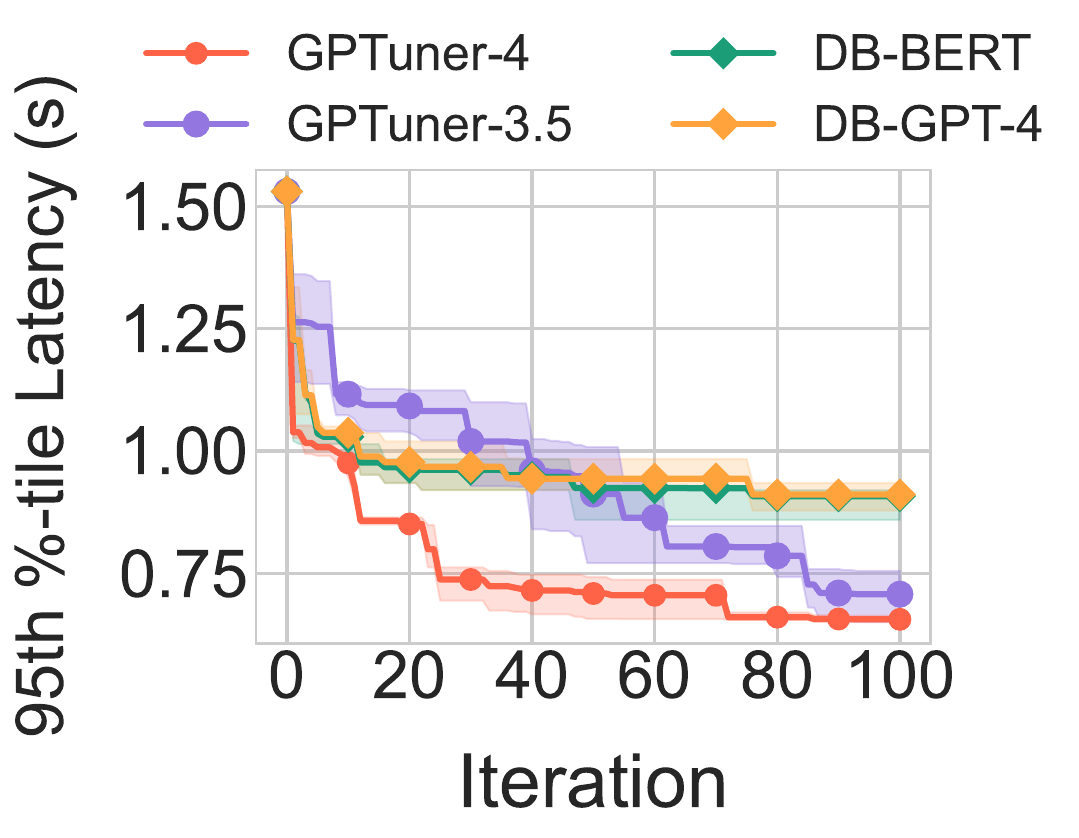}
             
            \caption{Effect of Language Models (bottom-left is better)}
            \label{fig: language-models}
        \end{figure}
    \end{minipage}
     
\end{figure*}

\color{black}
\subsection{Scalability Study}
\noindent \textbf{Database Size.} We study the impact of database size on tuning performance by varying the scale factor of TPC-H from 1 to 10 and 50. As shown in Figure \ref{fig: effect-data-size}, compared to other methods, \system finds better configurations in much fewer iterations in all sizes. When the factor is 50, \system identifies a configuration better than any other methods just in the 30-th iteration and finally achieves a 42.5\% reduction in latency. An interesting observation is that knowledge-enhanced approaches (i.e., \textsc{GPTuner} and DB-BERT) are affected by the increasing of database sizes slightly, while other methods suffer from it. This can be attributed to the fact that the complexity of modeling the relation between configurations and DBMS performance is higher in larger sizes, since more performance bottlenecks are revealed. \system learns such experience directly from domain knowledge rather than through iterative trial and error, and thus showcases superior performance, a result similar to that in \cite{10.1145/3514221.3517843}.


\noindent \textbf{Search Space Dimensionality.} We study the impact of space dimensionality on tuning performance by varying the number of target knobs from 50 to 100 and 150. Note that DB-BERT is excluded in this experiment since it relies on a frequency-based selection strategy to tune only a fixed set of knobs mentioned in the input documents (e.g., it tunes 22 knobs when tuning TPC-H for PostgreSQL), making it infeasible to manually control the number of target knobs \cite{10.1145/3514221.3517843}. As shown in Figure \ref{fig: effect-space-dimension}, \textsc{GPTuner} consistently showcases the best performance in all space sizes. While other approaches perform well in low-dimensional case, their performance deteriorates in high-dimensional cases. This is because although the space dimensionality is fixed for all methods, \system still benefits from its Range Optimization as discussed in Section \ref{subsec: space-reduction}.

     
     

\begin{table*}[t]
\centering
\caption{\color{black}Initial Profiling Overheads Statistics\color{black}}
\label{table: overheads}
 
\small
\setlength{\tabcolsep}{2pt}
\renewcommand{\arraystretch}{1}
\begin{tabular}{|cccccccccccccccccccc|}
\hline
\multicolumn{1}{|c|}{\multirow{3}{*}{Complex.}} & \multicolumn{1}{c|}{\multirow{3}{*}{Knob \#n}} & \multicolumn{9}{c|}{Tuning Lake}                                                                                                                                                                                                                                                                   & \multicolumn{9}{c|}{Structured Knowledge}                                                                                                                                                                                                                                      \\ \cline{3-20} 
\multicolumn{1}{|c|}{}                            & \multicolumn{1}{c|}{}                               & \multicolumn{3}{c|}{Token (k)}                                                                 & \multicolumn{3}{c|}{Money (USD)}                                                               & \multicolumn{3}{c|}{Time (s)}                                                                    & \multicolumn{3}{c|}{Token (k)}                                                                  & \multicolumn{3}{c|}{Money (USD)}                                                               & \multicolumn{3}{c|}{Time (s)}                                               \\ \cline{3-20} 
\multicolumn{1}{|c|}{}                            & \multicolumn{1}{c|}{}                               & \multicolumn{1}{c|}{3.5} & \multicolumn{1}{c|}{4} & \multicolumn{1}{c|}{4-turbo} & \multicolumn{1}{c|}{3.5} & \multicolumn{1}{c|}{4} & \multicolumn{1}{c|}{4-turbo} & \multicolumn{1}{c|}{3.5} & \multicolumn{1}{c|}{4}   & \multicolumn{1}{c|}{4-turbo} & \multicolumn{1}{c|}{3.5} & \multicolumn{1}{c|}{4}  & \multicolumn{1}{c|}{4-turbo} & \multicolumn{1}{c|}{3.5} & \multicolumn{1}{c|}{4} & \multicolumn{1}{c|}{4-turbo} & \multicolumn{1}{c|}{3.5} & \multicolumn{1}{c|}{4}   & 4-turbo \\ \hline
\multicolumn{1}{|c|}{\multirow{4}{*}{O (n)}}      & \multicolumn{1}{c|}{50}                             & \multicolumn{1}{c|}{124.4}     & \multicolumn{1}{c|}{160.1} & \multicolumn{1}{c|}{148.8}       & \multicolumn{1}{c|}{0.2}       & \multicolumn{1}{c|}{6.0}   & \multicolumn{1}{c|}{2.1}         & \multicolumn{1}{c|}{833.0}     & \multicolumn{1}{c|}{4155.0}  & \multicolumn{1}{c|}{3023.0}      & \multicolumn{1}{c|}{558.0}     & \multicolumn{1}{c|}{578.0}  & \multicolumn{1}{c|}{558.4}       & \multicolumn{1}{c|}{0.9}       & \multicolumn{1}{c|}{18.7}  & \multicolumn{1}{c|}{5.9}         & \multicolumn{1}{c|}{1183.0}    & \multicolumn{1}{c|}{4888.0}  & 1805.0      \\ \cline{2-20} 
\multicolumn{1}{|c|}{}                            & \multicolumn{1}{c|}{100}                            & \multicolumn{1}{c|}{254.6}     & \multicolumn{1}{c|}{324.1} & \multicolumn{1}{c|}{300.4}       & \multicolumn{1}{c|}{0.4}       & \multicolumn{1}{c|}{12.2}  & \multicolumn{1}{c|}{4.2}         & \multicolumn{1}{c|}{1679.0}    & \multicolumn{1}{c|}{7939.0}  & \multicolumn{1}{c|}{6021.0}      & \multicolumn{1}{c|}{1121.0}    & \multicolumn{1}{c|}{1162.7} & \multicolumn{1}{c|}{1105.5}      & \multicolumn{1}{c|}{1.7}       & \multicolumn{1}{c|}{37.8}  & \multicolumn{1}{c|}{11.7}        & \multicolumn{1}{c|}{2371.0}    & \multicolumn{1}{c|}{9738.0}  & 4985.0      \\ \cline{2-20} 
\multicolumn{1}{|c|}{}                            & \multicolumn{1}{c|}{150}                            & \multicolumn{1}{c|}{379.5}     & \multicolumn{1}{c|}{482.7} & \multicolumn{1}{c|}{443.9}       & \multicolumn{1}{c|}{0.6}       & \multicolumn{1}{c|}{18.2}  & \multicolumn{1}{c|}{6.2}         & \multicolumn{1}{c|}{2450.0}    & \multicolumn{1}{c|}{11742.0} & \multicolumn{1}{c|}{8626.0}      & \multicolumn{1}{c|}{1690.9}    & \multicolumn{1}{c|}{1738.4} & \multicolumn{1}{c|}{1657.7}      & \multicolumn{1}{c|}{2.6}       & \multicolumn{1}{c|}{56.4}  & \multicolumn{1}{c|}{17.5}        & \multicolumn{1}{c|}{3574.0}    & \multicolumn{1}{c|}{13958.0} & 6729.0      \\ \cline{2-20} 
\multicolumn{1}{|c|}{}                            & \multicolumn{1}{c|}{average}                  & \multicolumn{1}{c|}{2.5}       & \multicolumn{1}{c|}{3.2}   & \multicolumn{1}{c|}{3.0}         & \multicolumn{1}{c|}{0.003}       & \multicolumn{1}{c|}{0.1}   & \multicolumn{1}{c|}{0.04}         & \multicolumn{1}{c|}{16.3}      & \multicolumn{1}{c|}{78.3}    & \multicolumn{1}{c|}{57.5}        & \multicolumn{1}{c|}{11.3}      & \multicolumn{1}{c|}{11.6}   & \multicolumn{1}{c|}{11.1}        & \multicolumn{1}{c|}{0.02}       & \multicolumn{1}{c|}{0.4}   & \multicolumn{1}{c|}{0.1}         & \multicolumn{1}{c|}{23.8}      & \multicolumn{1}{c|}{93.1}    & 44.9        \\ \hline
\end{tabular}
 
\end{table*}

\subsection{Robustness Study} \label{subsec: robust}

\color{black}

\noindent \textbf{Effect of Error Correction Mechanisms.} We measure the reliability of the two mechanisms with two manually prepared datasets.


\noindent \textit{(1) Evaluating step 2 as a filter of step 1.} Firstly, we utilize GPT-4 to generate domain knowledge for 150 knobs from PostgreSQL, and then conduct a survey with a team of five database experts to annotate the generated knowledge. Among all pieces of knowledge, 123 of them are consistent with the \textit{system\_view} of PostgreSQL and the remaining 27 are not. Next, we use GPT-4 to filter out noisy knowledge as discussed in Section \ref{subsec: knowledge-extraction}. GPT-4 performs well in this task by identifying 20 pieces of noisy knowledge out of 27 pieces with a recall rate of 74\%, and classifying 109 pieces of correct knowledge out of 123 pieces with a specificity rate of 88.6\%. In summary, equipped with our step 2 as a filter of noisy knowledge, we improve the accuracy of input knowledge from 82\% to 94\%.


\noindent \textit{(2) Evaluating step 4 as a corrector of step 3.} At first, we prepare domain knowledge from GPT-4, manuals and web for 150 knobs from PostgreSQL. Next, we use GPT-3.5 and GPT-4 to complete step 3 and step 4, where they result in 7 and 4 inconsistencies respectively and rewrite all of them correctly. Since the number of inconsistencies is low and we want to further evaluate the LLM's reliability, we additionally prepare a dataset containing 50 summaries that are factual inconsistent with the original knowledge. Among the 50 incorrect pieces of knowledge, GPT-4 detects 39 of them are incorrect with an accuracy of 78\%, and rewrite all of them correctly with an accuracy of 100\%. The degradation of accuracy can be attributed to the complexity and diversity of the manually prepared knowledge, which pose challenges for LLM to process. Given that GPT-4 detects and corrects all inconsistencies in the real dataset, it is feasible to utilize GPT-4 as a corrector of step 3.

\color{black}

\color{black}
\noindent \textbf{Effect of Domain Knowledge Quality.} We study how \textsc{GPTuner}'s performance is affected by the quality of input domain knowledge. 

\noindent \textit{(1) Outdated Knowledge.} We only use PostgreSQL manual of version 9.1 as the knowledge input for \system to tune PostgreSQL v14.9, and this version is denoted as ``GPTuner-outdated''. 

\noindent \textit{(2) Inaccurate Knowledge.} Firstly, we use the domain knowledge with an accuracy of 82\% and 94\% from the ``Evaluating step 2 as a filter of step 1'' part, which are denoted as ``GPTuner-82'' and ``GPTuner'' respectively. Next, we use the knowledge from the ``Evaluating step 4 as a corrector of step 3'' part. Specifically, among all the target knobs, we randomly select half of them and assign the factual inconsistent knowledge to them. This procedure is repeated for five times and we present two representative results, which are denoted as ``GPTuner-50-1'' and ``GPTuner-50-2'', respectively. 

As shown in Figure \ref{fig: knowledge-quality}, the higher the knowledge quality, the better the tuning performance. Note that outdated knowledge only affects tuning performance slightly, while inaccurate knowledge impacts tuning performance with different degrees. This is related to our motivation in Section \ref{subsec: knob-select}, highlighting that some knobs are crucial in determining DBMS performance and \system benefits a lot if the knowledge about these important knobs is correct. On one hand, in production environment, we encourage users to double check such knowledge. On the other hand, given that \system under the noisy knowledge inputs is still better than the best optimizer without knowledge input (i.e., SMAC), we believe \system is robust enough to learn from mistakes in its BO stage and achieve satisfactory performance optimization ultimately.

\color{black}

\noindent \textbf{Effect of Different Language Models.} We analyze the effect of LLM on tuning performance. Specifically, DB-BERT is upgraded to use GPT-4 (``DB-GPT-4''), and \textsc{GPTuner} is tested with ``gpt-3.5-turbo'' and ``gpt-4'' (``GPTuner-3.5'' and ``GPTuner-4''). As shown in Figure \ref{fig: language-models}, DB-BERT does not benefit from the improvement of language model, implying its design is less dependent on the quality of language model. Both versions of \system outperform DB-BERT methods, and \system significantly benefits from GPT-4, demonstrating its ability to leverage more sophisticated LLM.

\color{black}
\subsection{Cost Analysis} \label{subsec: cost-analysis}

There are two kinds of overheads in the context of DBMS knob tuning: (1) Initial Profiling Overhead, and (2) Runtime Overhead. Initial Profiling Overhead is the time required to collect training samples or to pre-train models before the real tuning process (e.g., OtterTune \cite{10.1145/3035918.3064029} demands ``over 30k trials per DBMS'' to collect training samples). Runtime Overhead is the time taken by an optimizer to suggest the next configuration to evaluate. In practice, runtime overhead (e.g., < 1 second for SMAC \cite{10.14778/3538598.3538604}) is much less than the DBMS evaluation time (e.g., minutes to hours), and thus recent works focus on \textit{sample efficiency} (i.e., the number of DBMS evaluation times it takes to reach a given level of performance improvement) rather than the optimization time \cite{10.14778/3551793.3551844, 10.14778/3538598.3538604}. Since \system introduces no extra runtime overhead compared to the BO-based optimizer, we only discuss the Initial Profiling Overhead next. 

Given a specific DBMS product (e.g., PostgreSQL and MySQL), we only need to conduct domain knowledge-related stages for one time. We can use the built results repeatedly in the future until there are major version updates for that DBMS, introducing a large number of changes of knob features (e.g., DBMS vendors extensively add new knobs, delete old knobs or even change the functions of existing knobs). We present the number of tokens consumed, the monetary costs and the time required using three language models under three different number of target knobs. As presented in Table \ref{table: overheads}, the computational complexity is proportional to the number of target knobs (i.e., $O(n)$ where n is the number of target knobs), and the costs scale almost linearly with the increasing of $n$. Without parallelism optimization, it takes ``gpt-4-turbo'' about 2101.6k tokens, 23.7 USD and 4 hours to finish the knowledge preparation and transformation for 150 knobs from PostgreSQL. Given that the built results are reusable, and the costs of \system are much lower than previous approaches (e.g., it takes more than three months to initialize the knowledge base for OtterTune and this base requires rebuilding from scratch for changes in hardware components and software versions \cite{10.1145/3035918.3064029, 10.14778/3457390.3457404}), we conclude that the costs introduced in \textsc{GPTuner} are feasible and practical.

\color{black}

\subsection{Ablation Study} \label{subsec: ablation}
We conduct a comprehensive ablation study to reveal the benefits of each module in \textsc{GPTuner}. We omit this study due to space limitation, please refer to our technical report \cite{technical_report} for more details.

\color{black}

\color{black}
\section{Conclusion and Outlook}

\color{black}
This paper introduces \textsc{GPTuner}, a manual-reading database tuning system that leverages domain knowledge to enhance the knob tuning process. Extensive experiments are conducted to demonstrate the effectiveness, scalability and robustness of \textsc{GPTuner}.
\color{black}

\noindent \textbf{Online Tuning.} We follow the tuning paradigm of most existing approaches by tuning a copy DBMS offline, and then deploy the satisfactory configuration on the DBMS in production environment. However, it takes considerable resources to make a copy DBMS for offline tuning. The lack of a practical online tuning paradigm comes from the fact that a configuration sampled from the large configuration space could make the system crash, and such insecurity is disastrous to online services. In the future, we will explore how to use domain knowledge to eliminate dangerous configurations.



\noindent \textbf{Optimizing other Data Processing Systems.} The idea of using domain knowledge to enhance black-box optimization algorithms is generic, and can be utilized to enhance other data processing systems like Spark \cite{10.1145/2934664} and Flink \cite{ApacheFS}. In the future, we will identify and solve the unique challenges when targeting at other systems.

\color{black}

%


\section*{Acknowledgements}
Jianguo Wang acknowledges the support of the National Science Foundation under Grant Number \href{https://www.nsf.gov/awardsearch/showAward?AWD_ID=2337806}{2337806}.

\balance   
\bibliographystyle{ACM-Reference-Format} 
\bibliography{paper}

\end{document}